\documentclass[useAMS,fleqn]{mnras}
\usepackage{times,amssymb,amsmath,color,verbatim,graphicx,url,float}

\title[KiDS-450: Cosmological Constraints from Weak Lensing Peak Statistics-I]
{KiDS-450: Cosmological Constraints from Weak Lensing Peak Statistics-I: 
Inference from Analytical Prediction of High Signal-to-Noise Ratio Convergence Peaks} 
\author[Shan et al.]{HuanYuan Shan$^{1}$\thanks{E-mail: shanhuany@gmail.com},
Xiangkun Liu$^{2,3}$\thanks{E-mail: liuxk@ynu.edu.cn},
Hendrik Hildebrandt$^{1}$, Chuzhong Pan$^{3}$, \newauthor
Nicolas Martinet$^{1}$, Zuhui Fan$^{3}$, Peter Schneider$^{1}$, Marika Asgari$^{4}$, \newauthor 
Joachim Harnois-D\'{e}raps$^{4}$, Henk Hoekstra$^{5}$, Angus Wright$^{1}$, J\"org P. Dietrich$^{6,7}$, \newauthor 
Thomas Erben$^{1}$, Fedor Getman$^{8}$, Aniello Grado$^{8}$, Catherine Heymans$^{4}$, Dominik Klaes$^{1}$, \newauthor 
Konrad Kuijken$^{9}$, Julian Merten$^{10,11}$, Emanuella Puddu$^{8}$, Mario Radovich$^{12}$, Qiao Wang$^{13}$\\
$^{1}$ Argelander-Institut f\"ur Astronomie, Auf dem H\"ugel 71, 53121 Bonn, Germany\\
$^{2}$ South-Western Institute for Astronomy Research, Yunnan University, Kunming 650500, China\\
$^{3}$ Department of Astronomy, School of Physics, Peking University, Beijing 100871, China\\ 
$^{4}$ Institute for Astronomy, University of Edinburgh, Royal Observatory, Blackford Hill,
Edinburgh EH9 3HJ, UK\\
$^{5}$ Leiden Observatory, Leiden University, P.O.Box 9513, 2300RA Leiden, The Netherlands\\
$^{6}$ Faculty of Physics, Ludwig-Maximilians-Universit\"at, Scheinerstr. 1, 81679 Munich, Germany\\
$^{7}$ Excellence Cluster Universe, Boltzmannstr.\ 2, 85748 Garching, Germany\\
$^{8}$ INAF - Astronomical Observatory of Capodimonte, Via Moiariello 16, 80131 Napoli, Italy\\
$^{9}$ Leiden University, P.O.Box 9513, 2300RA Leiden, The Netherlands\\
$^{10}$ Department of Physics, University of Oxford, Keble Road, Oxford OX1 3RH, UK\\
$^{11}$ INAF, Osservatorio Astronomico di Bologna, via Pietro Gobetti 93/3, 40129 Bologna, Italy\\
$^{12}$ INAF - Osservatorio Astronomico di Padova, via dell'Osservatorio 5, 35122 Padova, Italy\\
$^{13}$ Key laboratory for Computational Astrophysics, National Astronomical Observatories, Chinese Academy of Sciences, Beijing, 100012, China}

\begin{document}
\date{Accepted \dots . Received \dots; in original form \dots}

\label{firstpage}

\maketitle

\begin{abstract}

This paper is the first of a series of papers constraining cosmological parameters with weak
lensing peak statistics using $\sim 450~\rm deg^2$ of imaging data from the Kilo Degree Survey (KiDS-450).
We measure high signal-to-noise ratio (SNR: $\nu$) weak lensing convergence peaks in the range of $3<\nu<5$,  
and employ theoretical models to derive expected values. These models are validated using a suite of
simulations. We take into account two major systematic effects, the boost factor and the effect of 
baryons on the mass-concentration relation of dark matter haloes. In addition, we investigate the impacts 
of other potential astrophysical systematics including the projection effects of large scale 
structures, intrinsic galaxy alignments, as well as residual measurement uncertainties in the shear 
and redshift calibration. Assuming a flat $\Lambda$CDM model, we find constraints for
$S_{\rm 8}=\sigma_{\rm 8}(\Omega_{\rm m}/0.3)^{0.5}=0.746^{+0.046}_{-0.107}$ according to the degeneracy
direction of the cosmic shear analysis and
$\Sigma_{\rm 8}=\sigma_{\rm 8}(\Omega_{\rm m}/0.3)^{0.38}=0.696^{+0.048}_{-0.050}$ based on the derived 
degeneracy direction of our high-SNR peak statistics. The difference between the power index of 
$S_{\rm 8}$ and in $\Sigma_{\rm 8}$ indicates that combining cosmic shear with peak statistics has the 
potential to break the degeneracy in $\sigma_{\rm 8}$ and $\Omega_{\rm m}$. Our results are consistent with 
the cosmic shear tomographic correlation analysis of the same dataset and $\sim 2\sigma$ lower than 
the Planck 2016 results.

\end{abstract}


\section{Introduction}

Large scale structures (LSS) in the Universe produce coherent distortions on the image of background 
galaxies, an effect caused by weak gravitational lensing (WL) and generally known as cosmic shear. By measuring 
the shapes of these galaxies, we are able to extract information about the foreground matter distribution 
(Bartelmann \& Schneider 2001). This is an important cosmological probe, however the shear signals are 
very weak, typically a few percent. In order to be able to measure cosmological parameters, we need very accurate
shape measurements for a vast number of distant faint and small galaxies, which is extremely challenging. 
Tremendous efforts have been made in observational developments (e.g., Erben et al. 2013; Kuijken et al. 2015; 
Jarvis et al. 2016; Hildebrandt et al. 2016; de Jong et al. 2015, 2017; Aihara et al. 2017; 
Mandelbaum et al. 2017; Zuntz et al. 2017), and methodological advances in extracting shape 
information (e.g., Hoekstra et al. 2015; Mandelbaum et al. 2015; Fenech Conti et al. 2017) and in statistical
analysis (see Kilbinger et al. 2015 and references therein). These have proved the feasibility of using WL 
effects in cosmological studies. The results from recent large surveys, including the Canada-France-Hawaii 
Telescope Lensing Survey (CFHTLenS\footnote{\url{http://www.cfhtlens.org/}}; Heymans et al. 2012), the Kilo 
Degree Survey (KiDS\footnote{\url{http://kids.strw.leidenuniv.nl/}}; Hildebrandt et al 2017) and the Dark 
Energy Survey (DES\footnote{\url{http://www.darkenergysurvey.org/}}; Troxel et al. 2017), have further 
strengthened their important roles. With ongoing and next generation surveys, such as the 
Subaru Hyper SuprimeCam lensing survey (HSC\footnote{\url{http://hsc.mtk.nao.ac.jp/ssp/}}; Aihara et al. 2017), 
{\it Euclid}\footnote{\url{http://sci.esa.int/euclid/}} (Laureijs et al. 2011), 
the Large Synoptic Survey Telescope (LSST\footnote{\url{http://www.lsst.org/}}; Abell et al. 2009), WL will 
become one of the main cosmological probes, realizing that much tighter controls of systematics are necessary.

The recent cosmic shear two-point correlation functions (2PCFs) analysis using data from $450$ square degrees 
of the Kilo Degree Survey (in Hildebrandt et al 2017, KiDS-450 hereafter) found a $2.3\sigma$ tension on the 
value of $S_{\rm 8}=\sigma_{\rm 8}(\Omega_{\rm m}/0.3)^{0.5}$ in comparison with that expected from the cosmic 
microwave background (CMB) measurements of the Planck satellite (Planck Collaboration et al. 2016a). 
Here $\Omega_{\rm m}$ and $\sigma_{\rm 8}$ are, respectively, the present matter density in units of the 
critical density, and the root-mean-square (rms) of the linear density fluctuations smoothed on scale 
of $8h^{-1}\hbox{Mpc}$. The KiDS-450 constraints are in agreement with other cosmic shear studies 
(Heymans et al. 2013; Joudaki et al. 2017a; Troxel et al. 2017), galaxy-galaxy lensing (Leauthaud et al. 2017), 
and pre-Planck CMB constraints (Calabrese et al. 2017). Understanding such a tension is currently an important 
aspect of research in the field.

The typical mean redshift of source galaxies in current WL surveys is $z<1$, and thus the WL signal is
sensitive to late-time structure formation. On the other hand, the
CMB properties are primarily affected by physical processes at early times. The tension between the results
obtained from these two probes might indicate missing ingredients in our current cosmological model.
To answer this, however, we need to first scrutinise carefully whether the tension arises unphysically from 
residual systematic errors in the analysis of different probes. For WL probes, different statistical quantities 
can respond differently to systematics. Thus it is helpful to perform cosmological studies with same WL
data, but using different statistical analyses. In this paper, we perform a WL peak analysis using the 
KiDS-450 data, derive an independent measurement of $S_{\rm 8}$ and compare our results with the cosmic shear 
results obtained from Hildebrandt et al. (2017).

In WL cosmological studies, the cosmic shear two-point statistics are the most commonly used statistical tools 
in probing the nature of dark matter (DM) and the origin of the current accelerating expansion of the Universe 
(e.g., Kilbinger et al. 2013; Heymans et al. 2013; Jarvis et al. 2016; Jee et al. 2016; Joudaki et al. 2017a;
Hildebrandt et al. 2017; Troxel et al. 2017). It is, however, insensitive to the non-Gaussian information 
encoded in nonlinear structure formation. WL peaks, on the other hand, are high signal regions, that are closely 
associated with massive structures along the line-of-sight (LOS). Their statistics is a simple and effective 
way to capture the non-Gaussian information in the WL field, and thus highly complementary to the cosmic shear 
2PCF (e.g., Kruse \& Schneider 1999; Dietrich \& Hartlap 2010; Shan et al. 2012, 2014; Marian et al. 2012, 2013; 
Lin \& Kilbinger 2015; Martinet et al. 2015; Hamana et al. 2015; Liu et al., 2015a, b, 2016;
Kacprzak et al. 2016). 

With recent wide-field WL imaging surveys, several measurements of WL peak counts have been performed, and
subsequent cosmological constraints have been derived. With the shear catalogue (Miller et al. 2013)
from CFHTLenS, Liu et al. (2015a)
generated convergence maps with various Gaussian smoothing scales, and identified peaks from the maps as
local maxima. Based on interpolations from a set of simulation templates with varying
cosmological parameters of $(\Omega_{\rm m}, \sigma_{\rm 8}, w)$, constraints on these
were obtained. Combining WL peak counts with the convergence power spectrum, they found that the
constraints can be improved by a factor of about $2$. Considering the high-SNR peaks in
the Canada-France-Hawaii Telescope Stripe 82 survey (CS82), Liu et al. (2015b) derived constraints on cosmological
parameters $(\Omega_{\rm m}, \sigma_{\rm 8})$ using the theoretical model of Fan et al. (2010). With the same
method, Liu et al. (2016b) presented constraints on the f(R) theory with the CFHTLenS data.
Kacprzak et al. (2016) measured the shear peaks using aperture mass maps (Schneider 1996;
Bartelmann \& Schneider 2001) from the Dark Energy Survey Science Verification (DES-SV) data. To derive
cosmological constraints, they also adopted the simulation approach to produce WL
maps (Dietrich \& Hartlap 2010) spanning the $(\Omega_{\rm m}, \sigma_{\rm 8})$ plane. 

Compared to cosmological studies with clusters of galaxies (Vikhlinin et al. 2009; Rozo et al. 2010; 
Planck Collaboration 2016b), WL peak statistics can provide cosmological constraints that are free from 
potential selection effects (Angulo et al. 2012) and biases associated with cluster mass estimates.

The correspondence between WL peaks and DM haloes is not one-to-one. Indeed, most of the low signal-to-noise 
ratio (SNR) peaks are usually not associated with a dominant halo, and are instead generated by the 
projection of LSS along the LOS. Even for high-SNR peaks where the correspondence with massive haloes is 
clearly seen, many systematic effects, such as the shape noise contamination from the intrinsic ellipticities of 
source galaxies, the boost factor due to the member contamination and the blending in cluster regions, baryonic 
effects, the projection effects of LSS, and intrinsic alignments (IA), can complicate WL peak 
analysis (Tang \& Fan 2005; Yang et al. 2011, 2013; Hamana et al. 2012; Fu \& Fan 2014; Osato et al. 2015; 
Kacprzak et al. 2016; Liu \& Haiman 2016; Yuan et al. 2017). These can generate non-halo-associated peaks and 
also alter the significance of the peaks from DM haloes, thus affecting WL peak statistics. Understanding and 
quantifying these effects is key to connect the observed peak signal to the underlying cosmology. 

There are different approaches to predict WL peak counts: (i) generating WL simulation
templates densely sampled in cosmological-parameter space (Dietrich \& Hartlap 2010; Liu et al. 2015a;
Kacprzak et al. 2016); (ii) theoretical modelling taking into account different systematic effects,
using either a pure Gaussian random field analysis (Maturi et al. 2010) or a halo model plus the Gaussian
random noise applicable to high-SNR peaks (Fan et al. 2010; Yuan et al. 2017); (iii) modelling a stochastic 
process to predict WL peak counts by producing lensing maps using a halo distribution from a theoretical 
halo mass function (Lin \& Kilbinger 2015). This is physically similar to the halo model.

In this work, we perform WL peak studies using the KiDS-450 data. To confront the tension on 
$S_{\rm 8}$ measurement, we derive an independent constraint on $S_{\rm 8}$ from the abundance of high-SNR peaks
adopting the analytical model of Fan et al. (2010), in which the dominant shape noise effects have been fully
taken into account. We further explore the potential systematics on WL peak statistics. We compare our
results with the ones derived from the tomographic cosmic shear measurement from Hildebrandt et al. (2017), 
as well as those from previous WL peak studies. We also observe a difference between the degeneracy direction
of $(\Omega_{\rm m}, \sigma_{\rm 8})$ in WL peak statistics and in cosmic shear analysis. Therefore, instead
of $S_{\rm 8}$, we use $\Sigma_{\rm 8}=\sigma_{\rm 8}(\Omega_{\rm m}/0.3)^{\alpha}$ and fit the
slope $\alpha$ to the data.

This is the first of a series of papers on cosmological constraints from WL peak statistics
using KiDS-450. In the subsequent paper, by comparing with simulation templates
from Dietrich \& Hartlap (2010), Martinet et al. (2017) derive constraints with shear peak
statistics identified from aperture mass maps. Because the projection effects of LSS are included
in the simulations, an independent measurement of the value of $S_{\rm 8}$ can be obtained with the low- and
medium-SNR peaks. The different physical origins of low- and high-SNR peaks
indicate different cosmological information embedded in the peak statistics of different ranges.
Furthermore, we expect that the systematics affect these two analysis in different ways.
Therefore the consistency between the results from the two studies indicate their robustness.

This paper is structured as follows: In Sect.~2, we describe the KiDS-450 dataset.
In Sect.~3, we present the procedures of WL peak analysis. In Sect.~4, we discuss the systematic
effects. In Sect.~5, we derive the cosmological constraints with WL peak counts. A summary and
discussion are given in Sect.~6.

\section{The KiDS-450 Data} \label{sec:KiDS450}

The ongoing Kilo Degree Survey (KiDS: de Jong et al. 2015; Kuijken et al. 2015), designed for WL studies,
is a $1350~\rm deg^2$ optical imaging survey in four bands $(u,g,r,i)$ with $5\sigma$ limiting magnitudes of
$24.3, 25.1, 24.9, 23.8$, respectively, using the OmegaCAM CCD camera mounted at the
Cassegrain focus of the VLT Survey Telescope (VST).

In this paper, we use the KiDS-450 shear catalogue (Hildebrandt et al. 2017; de Jong et al. 2017), which consists
of $454$ tiles covering a total area of $449.7~\rm deg^2$. After excluding the masked regions, the effective 
survey area is $360.3~\rm deg^2$. The lensing measurements are performed on the $r$-band images with median 
seeing $0.66~\rm arcsec$. The KiDS-450 $r$-band images are processed with the \textsc{Theli} pipeline, which 
has been optimised for lensing applications (Erben et al. 2009, 2013). As the observing
strategy of the KiDS survey was motivated to cover the Galaxy And Mass Assembly (GAMA) fields
(Liske et al. 2015), the KiDS-450 dataset contains five patches (G9, G12, G15, G23, GS),
covering $(45.95, 91.96, 89.60, 81.61, 51.16)~\rm deg^2$, respectively.

Photometric redshifts (photo-z) $z_{\rm B}$ are derived using the Bayesian point estimates from \textsc{BPZ} 
(Benitez 2000; Hildebrandt et al. 2012). The source redshift distribution $n(z)$ is calculated through a 
weighted direct calibration technique based on the overlap with deep spectroscopic 
surveys (the so-called `DIR' method; Hildebrandt et al. 2017).

The ellipticities of the galaxies are derived using a `self-calibrating' version of the shape measurement method
\textsc{lensfit} (Miller et al. 2013; Fenech Conti et al. 2017). The multiplicative
shear calibration bias, $m$, is obtained from image simulations with $\sim1\%$ error for galaxies with
$z_{\rm B}\leq0.9$. The additive shear calibration bias $c$ is estimated empirically from the data by averaging
galaxy ellipticities in the different patches and redshift bins.

In this paper, we first split the galaxy sample into four tomographic bins $z_{\rm B}=([0.1,0.3], [0.3,0.5], 
[0.5,0.7], [0.7,0.9])$ per patch as in Hilbedrandt et al. (2017), and apply shear calibration corrections
per tomographic bin and patch. The additive correction is done on individual galaxies, and the multiplicative
correction is performed statistically (see Eq.~8). Because of the low effective number density
$n_{\rm eff}\sim7.5~\rm gals/arcmin^2$ within $0.1<z_{\rm B}\leq0.9$ of KiDS-450, there are
only $\sim2~\rm gals/arcmin^2$ in each redshift bin. Such low number densities prevent us from
performing WL peak analysis tomographically at this stage. Therefore, after the correction, we then 
combine all the galaxies with $0.1<z_{\rm B}\leq0.9$ for WL peak count analysis.

\section{Weak lensing peak analysis}\label{sec:wlpeak}

\subsection{Theoretical basics}

The distortion of galaxy shapes by the gravitational lensing effect can be described by the Jacobian matrix
$\boldsymbol A$, which is given by (e.g., Bartelmann \& Schneider 2001)
\begin{equation}
\boldsymbol{A}=(1-\kappa)
\begin{pmatrix}
 1-g_1  & -g_2 \\
 -g_2 & 1+g_1
\end{pmatrix},
\label{JacobianMatrix}
\end{equation}
where $\boldsymbol{g}=\frac{\boldsymbol{\gamma}}{1-\kappa}$ is the reduced shear written in the 
complex form of $g_1+{\rm i}g_2$. The quantities $\boldsymbol{\gamma}$ and $\kappa$ are the complex lensing
shear and convergence, respectively. They can be calculated from the second derivatives of the lensing potential,
and thus $\boldsymbol{\gamma}$ and $\kappa$ are not independent quantities. The convergence $\kappa$ is
related to the projected matter density along the LOS scaled by a geometric
factor.

The observed lensing quantity is the complex ellipticity $\boldsymbol {\epsilon}$, which contains both the 
reduced shear and shape noise from the intrinsic galaxy ellipticity (Seitz \& Schneider 1997). In order to 
identify WL peaks, we need to relate the shear to the convergence, which involves a mass reconstruction 
algorithm. To reduce the noise from finite measurements of the shear, the observed ellipticities are 
regularised on a mesh and smoothed by a filter function. This results in an estimate of the smoothed field 
of the reduced shear $\boldsymbol{g}$. From that, the convergence field can be reconstructed with the nonlinear 
Kaiser-Squires (KS) inversion (Kaiser \& Squires 1993; Kaiser et al. 1995; Seitz \& Schneider 1995). We can 
then identify WL peaks, defined as local maxima in the two-dimensional convergence field. Their abundance 
contains important cosmological information that we analyze in this paper.

In our analysis, we construct the convergence map tile by tile. Each KiDS tile is $1~\rm deg^2$. In order 
to keep more effective area while excluding the problematic boundary, we extend each tile 
to $1.2\times1.2~\rm deg^2$ using data from neighboring tiles. The regular mesh in each convergence map 
contains $512\times512$ pixels with a pixel size of $\sim 0.14~\rm arcmin$. Then, the outermost $43$ 
pixels ($\sim 6~\rm arcmin$) along each side of the extended tile boundaries are excluded to suppress 
the boundary effects. Moreover, for area of this size, we expect an insignificant mass-sheet degeneracy 
contribution (Falco et al. 1985).

As described above, we smooth the pixelated ellipticity field with a Gaussian function,
\begin{equation}
W_{\theta_{\rm G}}(\boldsymbol{\theta})=\frac{1}{\pi\theta_{\rm G}^2}\exp{\left( -\frac{|\boldsymbol{\theta}|^2}{\theta_{\rm G}^2}\right )},
\label{window}
\end{equation}
where $\theta_{\rm G}$ is the smoothing scale. Hamana et al. (2004) found that 
$\theta_{\rm G}\sim1-2~\rm arcmin$ is an optimal choice for detecting massive haloes 
with $M\gtrsim10^{14}h^{-1}\hbox{M}_{\odot}$ at intermediate redshifts. In this paper, we 
take $\theta_{\rm G}=2~\rm arcmin$ so that $>30$ galaxies can be included in the
smoothing kernel effectively. Consequently, the Gaussian approximation for the shape noise field should be 
valid, according to the central limit theorem (Van Waerbeke 2000). The mean rms of the smoothed shape noise 
field is $\sigma_0\sim 0.023$, much larger than the contribution from the projection effect of 
LSS (discussed in Sect.~4.3), hence is dominant on our smoothed convergence maps.

\subsection{Weak lensing peak model}

In this work, we adopt a theoretical approach to derive the cosmological constraints from WL peak counts. 
Fan et al. (2010) presented a model taking into account the effects of shape noise, including 
the noise-induced bias and dispersion on the SNR of true peaks corresponding to massive DM haloes, the
spurious peaks induced by the shape noise of background sources, along with the enhancement of the pure
noise peaks near massive DM haloes. 

Specifically, this model assumes that the true high-SNR peaks are caused mainly by the existence of 
individual massive DM haloes (Hamana et al. 2004; Yang et al. 2011; Liu \& Haiman 2016) and that the residual 
shape noise field is approximately Gaussian. Accordingly, the smoothed convergence field can be written as
$\kappa^{\rm (S)}_{\rm n} = \kappa^{\rm (S)} + n^{\rm (S)}$, where $\kappa^{\rm (S)}$ represents the
true lensing convergence from individual massive haloes, and $n^{\rm (S)}$ is the residual Gaussian
shape noise. Assuming $\kappa^{\rm (S)}$ is known from the halo density profile, the field
$\kappa^{\rm (S)}_{\rm n}$ is therefore a Gaussian random field modulated by $\kappa^{\rm (S)}$. 
The peak count distribution can therefore be derived using Gaussian statistics, in which the
dependence on $\kappa^{\rm (S)}$ and its first and second derivatives
$\kappa^{\rm (S)}_{i}=\partial \kappa^{\rm (S)}/\partial x_i$ and
$\kappa^{\rm (S)}_{ij}=\partial^2 \kappa^{\rm (S)}/\partial x_i\partial x_j (i=1,2)$ of $\kappa^{\rm (S)}$
reflect the modulation effect of DM halos. The surface number density of convergence peaks can 
then be written as
\begin{equation}
n_{\mathrm{peak}}(\nu)d\nu=n_{\mathrm{peak}}^{\rm h}(\nu)d\nu+n_{\mathrm{peak}}^{\rm f}(\nu)d\nu,
\label{npeaktwoterm}
\end{equation}
where $\nu=\kappa/\sigma_0$ is the SNR of a peak, and $n_{\mathrm{peak}}^{\rm h}(\nu)$ and
$n_{\mathrm{peak}}^{\rm f}$ denote the number densities of WL peaks within halo regions (the virial radius) 
and those in the field regions outside, respectively. 

\subsubsection{Peaks in halo regions}

The peak count within halo regions, containing both the true peaks from the DM haloes and noise peaks 
therein, can be written as
\begin{equation}
 n_{\mathrm{peak}}^{\rm h}(\nu)=\int{{\rm d}z\frac{{\rm d}V(z)}{{\rm d}z\,{\rm d}\Omega}}\int_{M_{\rm lim}}{{\rm d}M\,n(M,z)\,f_{\rm p}(\nu,M,z)},
\label{npeakc}
\end{equation}
where ${\rm d}V(z)$ is the cosmological volume element at redshift $z$, ${\rm d}\Omega$ is the solid angle
element, $n(M,z)$ is the halo mass function, for which we adopt the function obtained by Watson et al.
(2013). Note that the model concerns high-SNR peaks, which are mainly due to a
single massive halo. We thus apply a lower mass limit $M_{\rm lim}$, and only haloes with mass $M>M_{\rm lim}$ 
contribute to the integration in Eq.~(4). From our investigation with mock data (Appendix~C), we find 
that: (1) a mass limit $M_{\rm lim}=10^{14}h^{-1}\hbox{M}_{\odot}$ for peaks with $\nu>3$ is a suitable choice 
which is also physically meaningful, as it corresponds to clusters of galaxies; (2) the input 
cosmological parameters can be well recovered, suggesting the impact of the uncertainties in the model 
ingredients, such as the halo mass function, are insignificant concerning the current study. 
The term $f_{\rm p}$ denotes the number of peaks within the virial radius 
of a DM halo, and is given by
\begin{equation}
f_{\rm p}(\nu,M,z)=\int_{0}^{\theta_{\mathrm{vir}}}{\rm d}\theta\hbox{ } (2\pi \theta)\hbox{ } \hat {n}^{\rm c}_{\mathrm{peak}}(\nu,\theta,M,z),
\end{equation}
where $\theta_{\mathrm{vir}}=R_{\rm vir}(M,z)/D_{\rm A}(z)$ is the angular virial radius, and $D_{\rm A}$ is 
the angular diameter distance to the DM halo. The physical virial radius $R_{\rm vir}$ is calculated by 
\begin{equation}
R_{\rm vir}(M,z)=\bigg[\frac{3M}{4\pi\rho(z)\Delta_{\rm vir}(z)}\bigg]^{1/3},
\label{virialr}
\end{equation}
where $\rho(z)$ is the background matter density of the Universe at redshift $z$ and the 
overdensity $\Delta_{\rm vir}$ is taken from Henry (2000). In our modeling, we limit the angular 
halo regions to $\theta_{\rm vir}$. The mass distributions outside it are regarded as parts of 
LSS contributions. Yuan et al. (2017) investigate in detail the LSS effects on peak statistics. 
For KiDS450, they are subdominant comparing to the impacts from shape noise.

The function 
$\hat {n}^{\rm c}_{\mathrm{peak}}(\nu,\theta,M,z)$ describes the surface number density of peaks at the 
location of $\theta$ from the centre of a halo, which can be derived using the theory of Gaussian random 
fields including the modulation effects from the DM halo contribution as follows
\begin{eqnarray}
&&\hat n^{\rm c}_{\mathrm{peak}}(\nu,\theta,M,z)=\exp \bigg [-\frac{(\kappa^{\rm (S)}_1)^2+(\kappa^{\rm (S)}_2)^2}{\sigma_1^2}\bigg ]\nonumber \\
&&\times \bigg [ \frac{1}{2\pi\theta_*^2}\frac{1}{(2\pi)^{1/2}}\bigg ]
\exp\bigg [-\frac{1}{2}\bigg ( \nu-\frac{\kappa^{\rm (S)}}{\sigma_0}\bigg )^2\bigg ] \nonumber \\
&&\times \int_0^{\infty} {\rm d}x_N\bigg \{ \frac{1}{ [2\pi(1-\gamma_N^2)]^{1/2}}\nonumber \\
&&\times \exp\bigg [-\frac{ [{x_N+(\kappa^{\rm (S)}_{11}+\kappa^{\rm (S)}_{22})/ \sigma_2
-\gamma_N(\nu_0-\kappa^{\rm (S)}/\sigma_0)}]^2}{ 2(1-\gamma_N^2)}\bigg ] \nonumber \\
&& \times  F(x_N)\bigg \},
\label{nchat}
\end{eqnarray}
with
\begin{eqnarray}
&&F(x_N)= \exp\bigg [-\frac{(\kappa^{\rm (S)}_{11}-\kappa^{\rm (S)}_{22})^2}{\sigma_2^2}\bigg ]
\times \nonumber \\
&& \int_0^{1/2}{\rm d}e_N \hbox{ }8(x_N^2e_N)x_N^2(1-4e_N^2) \exp(-4x_N^2e_N^2)
\times \nonumber \\
&& \int_0^{\pi} \frac{{\rm d}\theta_N}{ \pi} \hbox{ }
\exp\bigg [-4x_Ne_N\cos (2\theta_N)\frac{(\kappa^{\rm (S)}_{11}-\kappa^{\rm (S)}_{22})}{\sigma_2}\bigg ]. \nonumber \\
\label{fxn}
\end{eqnarray}
where $\theta_*^2=2\sigma_1^2/\sigma_2^2$, $\gamma_N=\sigma_1^2/(\sigma_0\sigma_2)$. The quantities 
$\sigma_i$ are the moments of the noise field $n^{\rm (S)}$ given by (e.g. Van Waerbeke 2000)
\begin{equation}
\sigma_i^2=\int {\rm d}k \hbox{ }k^{2i}\langle |\widetilde{n}^{\rm (S)}(k)|^2\rangle,
\label{noisesigma}
\end{equation}
where $\widetilde{n}^{\rm (S)}(k)$ is the Fourier transform of the noise field $n^{\rm (S)}$. 

For the density profile of dark matter halos, we adopt the Navarro-Frenk-White (NFW) 
distribution (Navarro et al. 1996, 1997):
\begin{equation}
\rho_{\rm NFW}(r)=\frac{\rho_{\rm s}}{(r/r_{\rm s})(1+r/r_{\rm s})^2},
\end{equation}
where $\rho_{\rm s}$ and $r_{\rm s}$ are the characteristic mass density and scale of a dark matter halo. 
The corresponding convergence $\kappa$ is obtained by integrating to the infinity along the LOS. 
We then smooth with the Gaussian function $W_{\theta_{\rm G}}$ (Eq.~2) to calculate the halo 
terms $\kappa^{\rm (S)}$, $\kappa^{\rm (S)}_i$, and $\kappa^{\rm (S)}_{ij}$.

We note that in WL analyses, there is not a consensus about the range of LOS integration for an NFW halo. 
We evaluate the impact of different LOS truncations on the peak analyses taking the models from 
Oguri \& Hamana (2011). It is found that their effects on our considered peak numbers are all well 
within $1\sigma$ statistical fluctuations.

The mass-concentration relation given in Duffy et al. (2008) is adopted in the calculation. 
In our fiducial analyses, the amplitude of the mass-concentration relation is considered as a free parameter to 
be fitted by the data simultaneously with cosmological parameters. 

For the redshift distribution of source galaxies, we take the DIR redshift distribution of KiDS-450 
data in the fiducial analysis but also consider other cases to test for the effect of redshift uncertainties. 
The impact of the uncertainties in the source redshift distribution on the measured WL peak counts is estimated 
from $200$ bootstrap resamples drawn from the full spectroscopic
redshift training catalogue (Hildebrandt et al. 2017). By analyzing different $n(z)$ distributions with
the same pipeline, we find that our peak analysis is essentially unaffected within the redshift
uncertainties. A similar conclusion is found in the cosmic shear analysis of Hildebrandt et al. (2017).

\subsubsection{Peaks in the field regions}

The density of pure noise peaks in the field region away from DM haloes is given by
\begin{equation}
 \begin{split}
n_{\mathrm{peak}}^{\rm f}(\nu)=\frac{1}{{\rm d}\Omega}\Big\{n_{\mathrm{ran}}(\nu)\Big[{\rm d}\Omega-\int {\rm d}z\frac{{\rm d}V(z)}{{\rm d}z}\\
   \times\int_{M_{\rm lim}} {\rm d}M\,n(M,z)\,(\pi \theta_{\rm vir}^{2})\Big]\Big\},
\end{split}
\label{npeakn}
\end{equation}
where $n_{\mathrm{ran}}(\nu)$ is the surface number density of pure noise peaks without foreground DM haloes.
It can be calculated with $\kappa^{\rm (S)}=0$, $\kappa^{\rm (S)}_i=0$, and $\kappa^{\rm (S)}_{ij}=0$. \\

We can see that, in the model, the cosmological information comes from the halo mass function, the internal
density profile of DM haloes, and the cosmological distances in the lensing efficiency factor as well as the cosmic
volume element. This model has been tested extensively with simulations (Fan et al. 2010; Liu et al. 2014). In
Appendix~A, we further test the model performance with the simulations from Dietrich \& Hartlap (2010) with
different underlying cosmological parameters, and has already been applied to derive cosmological constraints 
with observed WL peaks of CS82 and CFHTLenS data.

\subsection{Map making}\label{sec: map make}

In this section, we present the map making procedur from the KiDS-450 shear catalog. In order to build a
reliable WL peak catalog, three kinds of maps need to be generated for each tile.

(1) {\it Convergence map.}
Using the observed shear catalogue of KiDS-450, the smoothed shear field at positions $\boldsymbol \theta$ can 
be calculated by taking into account the multiplicative and additive calibration corrections.
\begin{equation}
\mathbf{\langle{\boldsymbol \epsilon}}_i\mathbf{\rangle(\boldsymbol{\theta})}=\frac { \sum_{j}W_{\theta_{\rm G}}(\boldsymbol{\theta_j}-\boldsymbol{\theta}) w(\boldsymbol{\theta_j}) \mathbf{\epsilon}^c_i(\boldsymbol{\theta_j})}{\sum_{j}W_{\theta_{\rm G}}(\boldsymbol{\theta_j}-\boldsymbol{\theta})w(\boldsymbol{\theta_j})(1+m_j)},
\label{smoothing}
\end{equation}
where $W_{\theta_{\rm G}}$ is the Gaussian smoothing function in Eq.~(2) with the smoothing scale
$\theta_{\rm G}=2~\rm arcmin$, $\epsilon_i^c=\epsilon_i-c_i$, where $\epsilon_i$ and $\epsilon_i^c$ are the
uncorrected and corrected ellipticity components, $m$ and $(c_1, c_2)$ are the multiplicative
and the additive bias corrections, respectively, and $w$ is the \textsc{lensfit} weight of source galaxy shape
measurements. The summation is over galaxis $j$ at  positions $\boldsymbol \theta_j$.

For the KiDS-450 lensing data with redshift $0.1<z_{\rm B}\leq0.9$, the average multiplicative and 
additive biases $(m,c)$ are quite small with $(\sim1.4\times 10^{-2}, \sim3.9\times 10^{-4})$,
respectively. Given that the residual uncertainty in the bias estimation is only $1\%$, it 
can only influence the theoretical predictions for peak counts with $\nu>3$
by $\sim1-2\%$. This is well within the statistical uncertainties of our measurement.

The additive bias, $c$, is obtained empirically from the data by averaging the measured ellipticities
in different KiDS patches and redshift bins. Their uncertainties are at the level of $\sim 6\times 10^{-5}$.
As discussed in Kacprzak et al. (2016), the additive bias systematics can vanish within the smoothing scale
except for the galaxies at the edges of survey masks. With the filling factor cut in our peak
analysis (see below), we expect a negligible impact of the additive bias on our results.

With the smoothed shear fields, the convergence map can be reconstructed iteratively for each individual tile
using the nonlinear KS inversion (Seitz \& Schneider 1995; Liu et al. 2014). Assuming $\kappa^{(0)}=0$ in a tile,
we have $\gamma^{(0)}=\langle \boldsymbol {\epsilon} \rangle$. At the $n-$th step, we can obtain $\kappa^{(n)}$ 
from $\gamma^{(n-1)}$. We then update $\gamma$ to
$\gamma^{(n)}=(1-\kappa^{(n)})\langle \boldsymbol {\epsilon} \rangle$ for the next iteration.
The reconstruction process is stopped when the converging accuracy of $10^{-6}$, defined to be the maximum
difference of the reconstructed convergence between the two sequential iterations, is reached.

(2) {\it Noise map.}
To estimate the shape noise properties in each tile, the $m$-corrected ellipticity of each
source galaxy is rotated by a random angle to destroy the lensing signal. Then following the same
reconstruction procedures described above in (1), we can obtain the convergence noise field for each
tile in KiDS-450.

(3) {\it Filling factor map.}
Because mask effects can influence the WL peak counts significantly (Liu et al. 2014), the regions
around masks should be excluded in the WL peak analysis. For that, we need to construct filling-factor maps
from the positions and weights of source galaxies. The filling factor is defined as the ratio of the true
source galaxy density to that of the randomly populated galaxy distribution as follows
\begin{equation}
f(\boldsymbol{\theta})=\frac{\sum_{j}W_{\theta_{\rm G}}(\boldsymbol{\theta_j}-\boldsymbol{\theta})w(\boldsymbol{\theta_j})}{\langle \sum_{n}W_{\theta_{\rm G}}(\boldsymbol{\theta_n}-\boldsymbol{\theta})
\tilde {w}(\boldsymbol{\theta_n})\rangle}.
\label{filling}
\end{equation}
Here the numerator is calculated from the observed galaxy positions $\boldsymbol{\theta_j}$ and weights
$w(\boldsymbol{\theta_j})$. The denominator is calculated by randomly populating galaxies over the full
area of an extended tile. Specifically, we first find for each tile the average number density of galaxies
in the area excluding the masked regions. We then randomly populate galaxies over the full field of the
extended tile including the masked regions. Each galaxy is then assigned a weight $\tilde {w}$ randomly
according to the weight distribution of the source galaxies. From this random galaxy distribution, we obtain
the denominator where the summation is over all galaxies.

With the filling factor maps, we can then identify and exclude regions around masks in the reconstructed
convergence maps for peak counting. To control the systematic effects from the masks, we remove the regions 
with filling factor values $f<0.6$ in the peak counting (Liu et al. 2014).

\subsection{Peak identification}

In a reconstructed convergence map, a peak is identified if its pixel value is higher than that of 
the $8$ neighbouring pixels.

We exclude a tile entirely if its effective galaxy number density $n_{\rm eff}<5.5~\rm arcmin^2$ to ensure the 
validity of the Gaussian noise and the approximate uniformity of the noise field (Appendix~B). After further 
rejecting the tiles that fail the filling factor requirement, the total area for the peak analysis 
is $\sim 304.2~\rm deg^2$.

We then divide peaks into different bins based on their SNR $\nu=\kappa/\sigma_0$, where $\sigma_0$ is the
mean rms of the noise estimated from the noise maps considering only the regions that passed all requirements. 
With $\theta_G=2~\rm arcmin$, we have $\sigma_0\sim0.023$. Due to limitations in the model, we only consider
peaks with $\nu>3$, corresponding to a smoothed $\kappa\gtrsim 0.07$. For higher SNR, we include those bins
that contain at least $10$ peaks to avoid the bias resulting from the large Poisson fluctuations. We thus
concentrate on the peaks in the range of $3<\nu<5$.

\subsection{Fitting method}

We use the model described in Sect.~3.2 to derive cosmological constraints from the observed WL peaks identified
from the convergence maps. We divide the measurements in four equally wide SNR 
bins $([3.0,3.5],[3.5,4.0],[4.0,4.5],[4.5,5.0])$ where the number of peaks in the last bin being $\sim 10$ and 
significantly larger in the other bins. We define the following $\chi^2$ to be minimised for cosmological parameter constraints,
\begin{equation}
 \chi_{p}^{2}=\sum_{i,j=1}^{4}\Delta N_{i}^{(p)}(\widehat{C_{ij}^{-1}})\Delta N_{j}^{(p)},
\label{chi2}
\end{equation}
where $\Delta N_i^{(p)}=N_{\rm peak}^{(p)}(\nu_i)- N_{\rm peak}^{(d)}(\nu_i)$ is the difference between the
theoretical prediction with cosmological model $p$ and the observed peak counts. 
The covariance matrix $C_{ij}$ is estimated from bootstrap analysis by resampling the $454$ tiles from 
the KiDS-450 data, and is given by
\begin{equation}
 C_{ij}=\frac{1}{R-1}\sum_{r=1}^{R}[N^r_{\rm peak}(\nu_i)-{N}_{\rm peak}^{(d)}(\nu_i)][N^r_{\rm peak}(\nu_j)-{N}_{\rm peak}^{(d)}(\nu_j)].
\label{covar}
\end{equation}
Here, $r$ denotes different bootstrap samples with the total number $R=10000$, and $N^r_{\rm peak}(\nu_i)$
is the peak count in the bin centred on $\nu_i$ from sample $r$. The unbiased inverse of the
covariance matrix can be then estimated as (Hartlap et al. 2007)
\begin{equation}
\widehat{\boldsymbol{C}^{-1}}=\frac{R-N_{\mathrm{bin}}-2}{R-1}(\boldsymbol{C}^{-1}),~~N_{\mathrm{bin}}<R-2
\label{nobiascov}
\end{equation}
where $N_{\mathrm{bin}}$ is the number of bins used for peak counting. In our analysis, we adopt the
bootstrap covariance estimated from the KiDS-450 data. Liu et al. (2015b) found that the
differences between the results from simulation sets and from bootstrap resampling are generally less 
than $10\%$ for the diagonal elements of the inverse.

With $N_{\mathrm{bin}}=4$, in this paper, we consider constraints on the most lensing-sensitive parameters
$(\Omega_{\rm m}, \sigma_{\rm 8})$ under the flat $\Lambda$CDM assumption. In our fiducial analysis,
the other parameters including the Hubble constant $h$, the power index of the initial density perturbation
spectrum $n_{\rm s}$ and the present baryonic matter density $\Omega_{\rm b}$ are fixed to
$h=0.7$, $n_{\rm s}=0.96$ and $\Omega_{\rm b}=0.046$. We also consider cases with different Hubble constant
to see if this uncertainty can affect the results significantly. Our Markov Chain Monte Carlo (MCMC)
fitting uses \textsc{CosmoMC} (Lewis \& Bridle 2002) modified to include the likelihood of WL
peak counts. We adopt flat priors in the range of $[0.05, 0.95]$ and $[0.2, 1.6]$
for $\Omega_{\rm m}$ and $\sigma_{\rm 8}$, respectively.

In Appendix~A, we further test the model performance by comparing with simulations from Dietrich \& Hartlap
(2010) of different $(\Omega_{\rm m}, \sigma_{\rm 8})$. In Appendix~C, we analyse KiDS-450-like mock data
based on our own simulations using the full peak analysis pipeline. It is shown that the derived
constraints from the mock data can recover the input cosmological parameters very well. 

\section{Systematics}\label{sec: systematics}

As discussed in previous sections, the measurement systematics, including the shear measurement bias and
photo-z errors, are negligible for our KiDS-450 WL peak analysis. However, we need to further understand 
the impact of astrophysical systematic effects, such as the boost factor due to cluster member contamination 
and the blending in cluster regions, baryonic effects, the projection effects of LSS, and intrinsic alignments 
of galaxies (IA).

\subsection{Boost factor}

The true high-SNR peaks that we detect are mainly due to individual massive clusters. Cluster member 
contamination to the source galaxy catalogue can however dilute the lensing signals (e.g., Mandelbaum et al. 2006; 
Miyatake et al. 2015; Dvornik et al. 2017). In addition, the galaxies in cluster regions can be 
blended because of galaxy concentration, resulting in lower shear measurement weights. Both these effects need 
to be accounted as a `boost factor' (Kacprzak et al. 2016).

With DES-SV data, Kacprzak et al. (2016) find that the boost factor correction is $<5\%$ for their shear peak
studies: the dilution of the signal by cluster member galaxies is minimal $(<2\%)$, and the effect of background
galaxies lost because of blending is $\sim5\%$ in the SNR of the highest-SNR peaks with $3.66\leq \nu \leq 4.0$ 
with aperture radius $\theta_{\rm max}=20~\rm arcmin$. We note that our peak analysis is different from that of
Kacprzak et al. (2016) (convergence vs. shear peaks, and Gaussian filter vs. NFW-like filter). The 
modelling of the cosmological dependence is also different (theoretical vs. simulation templates). Thus
the estimate of the boost factor of Kacprzak et al. (2016) may not be directly applicable here. In this 
section, we estimate the boost effect based on our analysis, drawing out the different conclusions to 
Kacprzak et al. (2016).

The boost factor effect on peak statistics results from the excess galaxy number density (filling factor) of
source galaxies near massive clusters, compared to the average number density. To estimate these 
differences, it is better to analyse the source galaxies near known clusters in the field rather than around 
peaks because a considerable fraction of peaks are non-halo-associated.

In the galaxy-galaxy lensing measurement with KiDS and GAMA data, Dvornik et al. (2017) find that the member
contamination for GAMA groups can reach up to $\sim 30\%$ at $75~\hbox{kpc}/h$ and decreases on larger scales.
In our analysis, we use a Gaussian smoothing with $\theta_{\rm G}=2~\rm{arcmin}$. This corresponds to a scale
of $\sim 300~\hbox{kpc}/h$ at redshift $\sim 0.2-0.3$. Then a member contamination of
$\sim 10\%$ is expected. On the other hand, GAMA groups have a typical mass
of $10^{13}\hbox{M}_{\odot}/h$ (Dvornik et al. 2017), smaller compared to those responsible 
for the high-SNR peaks.

We therefore use the cluster candidates from Radovich et al. (2017) found in $114~\rm deg^2$ of KiDS
regions. The mass of the cluster candidates is estimated using the richness as a proxy (Anderson 2015). To assess
the boost factor effect due to the member contamination and the blending effect in cluster regions, similar to
Kacprzak et al. (2016), we analyse the filling factor of source galaxies near these cluster
candidates. Specifically, in accord with the high-SNR peak studies, we consider clusters with
mass $M>10^{14}h^{-1}\hbox{M}_{\odot}$. In Appendix~D, we quantify the impact of the boost factor effects
on both the signal and the noise level for WL peak counts from KiDS-450 data. They 
can affect the peak abundance by $\sim(-2.0\%, -6.0\%, -14.0\%, -27.0\%)$ on the four SNR bins 
$([3.0, 3.5], [3.5, 4.0], [4.0, 4.5], [4.5, 5.0])$ for the best-fit cosmology. We include the boost factor
effect in our fiducial analysis to derive cosmological parameters constraints (see Sect.~5).

\subsection{Baryonic effects}

Although baryonic matter is subdominant compared to DM, it is subject to complicated physical 
processes such as heating, cooling and feedback from stars and AGNs, all of which can have significant
influence on structure formation. For the WL peak analysis, the baryonic effect can be estimated by how it 
changes the DM distribution in haloes.

Using a simplified model for the cooling and condensation of baryons at the centres of DM haloes,
Yang et al. (2013) claim that there is a large increase in the number of high-SNR
peaks, but the effects on low-SNR peaks are quite small.

On the other hand, including the feedback of supernovae, stars and AGNs, Osato et al. (2015) find that
the feedback effects can effectively reduce the mass of small DM haloes, eventually reducing the number of 
low-SNR WL peaks. Because of the smaller impact of feedback on the massive DM haloes (Velliscig et al. 2014), the 
high-SNR peak number is not significantly changed. Osato et al. (2015) also show that the high-SNR peaks are 
almost unaffected once all the contributions from radiative cooling and the various feedbacks are 
included, because these effects can partially compensate each other. In fact, the baryonic effects are 
only expected to generate $1\%-2\%$ biases on the $(\Omega_{\rm m}, \sigma_{\rm 8})$ constraints from 
high-SNR peak analysis (Osato et al. 2015). 

Studies of the baryonic effects on WL peak statistics have 
not yet reached an agreement. This is mainly due to the different physical processes considered in the different 
analyses. Because the details of the baryonic physics are complicated and remain to be fully understood,
it would be highly valuable if we could obtain some constraints on them from observations
simultaneously with cosmological parameters. In addition, a self-calibrated method can also
reduce biases on cosmological parameter constraints arising from improper assumptions about the 
baryonic sector. In our theoretical modelling, the dependence of WL peak counts on baryonic effects 
is explicit. It is therefore possible for us to carry out studies including self calibration. 

For high-SNR WL peak counts, it is a reasonable assumption that baryonic effects show up through modifying 
the density distribution of DM haloes (Duffy et al. 2010; Mead et al. 2015). We therefore include some 
freedom in the halo mass-concentration relation. Specifically, we take the power-law form of the 
mass-concentration relation for NFW haloes,
\begin{equation}
c_{\rm vir}=\frac{A}{(1+z)^{0.7}}\left(\frac{M_{\rm vir}}{10^{14}h^{-1}{M_\odot}}\right)^{\beta},
\label{cm}
\end{equation}
where $A=5.72$ and $\beta=-0.081$ are given in Duffy et al. (2008). The redshift dependence
$(1+z)^{0.7}$ is taken to be consistent with simulation results (Duffy et al. 2008; Bhattacharya et al. 2013).
In order to quantify the possible baryonic effects on the density profiles and also the impact of the
uncertainties of the mass-concentration relation, we allow the amplitude $A$ to be a free parameter in 
our fiducial analysis. With a wide flat prior of $[0, 20]$, we then perform the simultaneous constraints on the 
cosmological and structural parameters $(\Omega_{\rm m}, \sigma_{\rm 8}, A)$ (see Sect.~5). Comparing with the 
prediction of DM-only simulations, the derived $A$ tends to be somewhat higher. But the current peak counts 
can hardly put any meaningful constraints on $A$.

\subsection{The projection effects of LSS}

Previous studies have shown that WL peaks of different
SNR originates from different sources (Yang et al. 2011; Liu \& Haiman 2016). While, high-SNR peaks
originate primarily from individual massive DM haloes (see Sect.~33), low SNR peaks often result from 
the cumulative contributions of the LSS along the LOS.

However, the projection effects of LSS affect the measurements of peaks for all SNR (Hoekstra 2001; 
Hoekstra et al. 2011). With the model of Fan et al. (2010), Yuan et al. (2017) investigate in detail the 
projection effects of LSS on high-SNR peaks, which shows that the ratio of $\sigma_{0,\rm {LSS}}^2/\sigma_0^2$ 
can give a rough estimate of the importance of LSS in comparison with that of the shape 
noise, where $\sigma_{0,\rm {LSS}}$ is the rms of the smoothed convergence field from LSS excluding the 
massive halo contributions, and $\sigma_0$ is the rms of the residual shape noise. The higher the 
redshift and the larger the density of source galaxies, the more important the effect of LSS. 
For KiDS-450, the number density is relatively low and thus the shape noise is large. The 
median redshift is also relatively low with $\sim 0.65$. In this case, 
$\sigma_{0,\rm {LSS}}^2/\sigma_0^2\sim (0.006/0.023)^2\sim 0.07$, and thus the LSS effect is much
lower than that of the shape noise. Furthermore, the effective area used in our
peak analysis is $\sim 300~{\rm deg}^2$, and the statistical errors of peak counts are relatively large.
We therefore expect minor impacts of LSS in our current analysis.

In fact, the projection effects of LSS are naturally included in the mock simulation data. The unbiased
results of the cosmological constraints from the mocks (Appendix~C) suggest that the LSS projection effects
are indeed negligible and the model that does not account for LSS projections still provide a good 
fit to the mock data. We note that for KiDS, with
the increase of the survey area, the statistical errors of peak counts will
decrease and the tolerable levels of systematic errors will also decrease. Thus the LSS effect may need to be
included in the peak modelling in future analysis (Yuan et al. 2017).

Moreover, by comparing with simulation templates, the low-SNR shear peaks from the projection effects of LSS 
are used to probe the cosmological information in Paper II.

\subsection{Intrinsic alignments}

The IA signal of galaxies contains important information on the formation and evolution of galaxies in 
their DM environment. For the cosmic shear 2PCF measurements, the IA effects can be divided into two components: 
the intrinsic ellipticity correlations (II) and shear-ellipticity correlations (GI). They can contaminate 
the cosmic shear analysis.

Fan (2007) studied the influence of IA on the convergence peak counts, by modelling it as additional
terms to the moments of the shape noise. The full noise variance in a convergence map can then be written
as $\sigma_0^2=\sigma_{\rm 0,ran}^2+\sigma_{\rm 0,corr}^2$, where $\sigma_{\rm 0,ran}$ is the noise
contributed from the randomly oriented intrinsic ellipticities of source galaxies,
and $\sigma_{\rm 0,corr}$ denotes the additional contribution from IA (see Eq.~23 in Fan 2007).
For the KiDS-450 data, we have $\sigma_{\rm 0,ran}^2=0.023^2=5.3\times 10^{-4}$ with a $2~\rm arcmin$ Gaussian
smoothing. We can estimate $\sigma_{\rm 0,corr}^2<3.07\times10^{-6}$ with the IA
amplitude $A_{\rm IA}=1.10\pm0.64$ from
the cosmic shear constraints (Hildebrandt et al. 2017), which is much smaller than $\sigma_{\rm 0,ran}^2$.

Apart from contributing to the noise variance, IA can also affect the peak signal estimates. 
If there is a contamination of cluster members to the source catalogue and these members are
intrinsically aligned to the centre, the estimated lensing signal would be biased.
Using a simple model of radial alignment of satellite galaxies with a certain misalignment angle consistent
with simulations, Kacprzak et al. (2016) estimated the IA influence on the SNR of
shear peaks with the aperture mass statistics. They find that the IA effects can be important for high-SNR 
shear peaks. For peaks with SNR $\nu>4.5$, the number of shear peaks can change by about $30\%$.

On the other hand, observationally, Chisari et al. (2014) find that the IA signals in stacked clusters of the
Sloan Digital Sky Survey (SDSS) `Stripe 82' in the redshift range $0.1<z<0.4$ are consistent with zero.
Using a large number of spectroscopic members of $91$ massive galaxy clusters with a median redshift
$z_{\rm med}\sim0.145$, Sif\'on et al. (2015) also find that the IA signal of cluster members is consistent with
zero for all scale, colour, luminosity, and cluster mass investigated. Because high-SNR peaks are mainly due to 
individual massive DM haloes hosting clusters of galaxies, these observational results may indicate negligible 
IA effects for high-SNR peak signal estimates.

We further note that for our analysis here, the number of peaks with $\rm {SNR}>4.5$ is about $10$, for which the
Poisson statistical uncertainty reaches $\sim 33\%$. For such large statistical fluctuations, we
do not expect the IA contamination to matter. \\

To summarize, the measurement systematics (shear measurement bias and photo-z errors) and
some astrophysical systematic effects (the projection effects of LSS, IA) are insignificant for our
cosmological studies using WL peaks from KiDS-450, and will be neglected. On the other hand, in our
fiducial studies, we include the boost effect, which we find to be significant. We also allow 
the amplitude of the halo mass-concentration relation to vary to account for possible baryonic effects.

\section{Cosmological constraints from KiDS-450 peak analysis}

In this section we present cosmological constraints derived from the KiDS-450 WL peak analysis, incorporating 
both the boost factor and baryonic effects as discussed in Sect.~4. 

\begin{figure}
\includegraphics[width=0.5\textwidth]{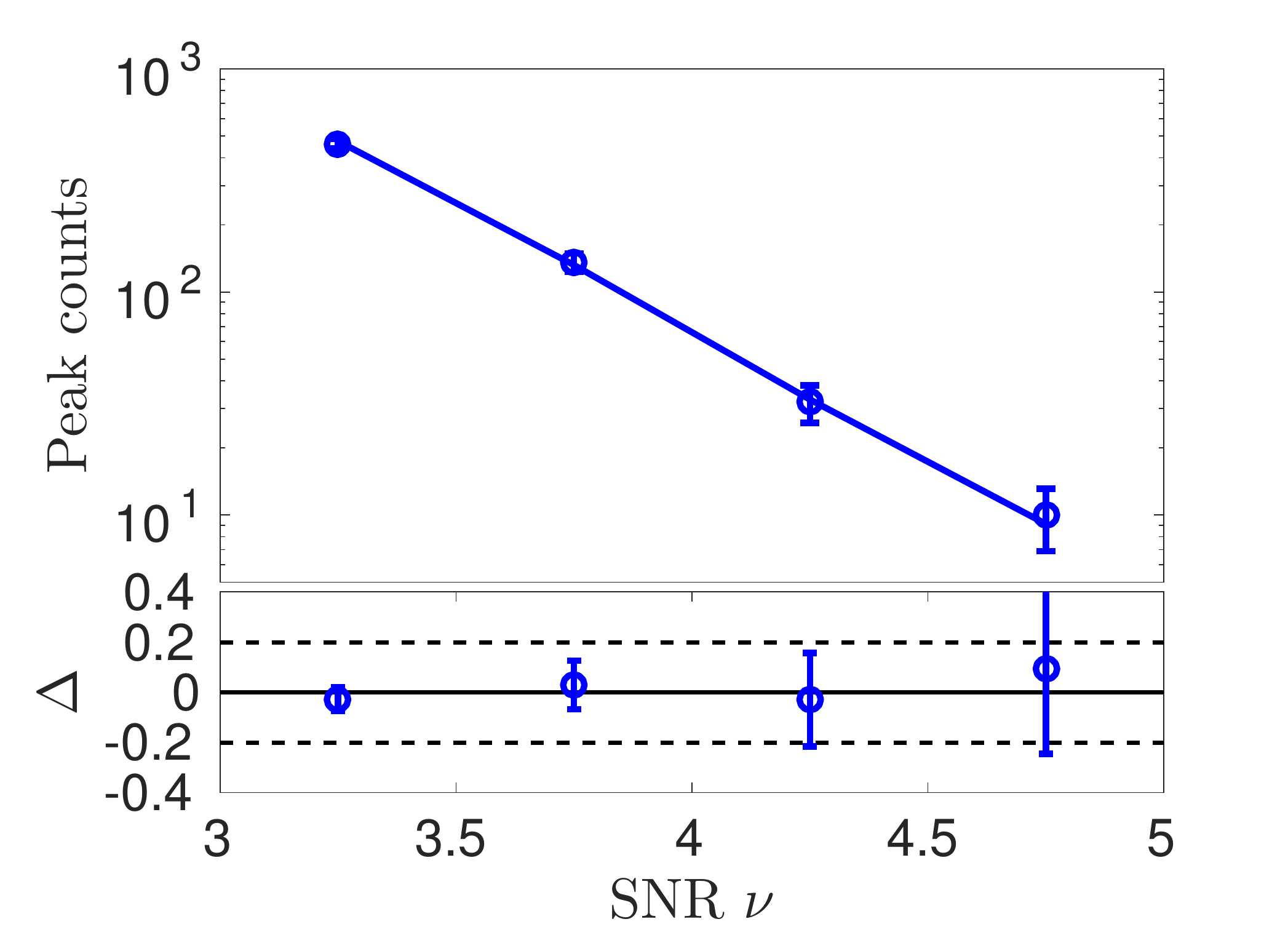}
\caption{Upper panel: The fiducial peak count distribution of the KiDS-450 data. The corresponding solid
line is the theoretical prediction with the best-fit cosmological parameters obtained from MCMC fitting. The
error bars are the square root of the diagonal terms of the covariance matrix.
Lower panel: The difference between the peak counts of the data and the best-fit theoretical predictions.}
\label{fig:kids450_peak}
\end{figure}

\begin{table}
\caption{The peak counts from the KiDS-450 data and of the theoretical predictions from the best-fit 
cosmological model in our fiducial analyses.}
\label{tab:peak}
\begin{center}
\begin{tabular}{ccc} \hline
   & $N^{\rm data}_{\rm peak}$ & $ N^{\rm fid}_{\rm peak}$ \\
\hline
$3.0\leq\nu<3.5$ & $462\pm23$ & $475.04$ \\
$3.5\leq\nu<4.0$ & $136\pm13$ & $132.11$ \\
$4.0\leq\nu<4.5$ & $32\pm6$ & $32.96$ \\
$4.5\leq\nu<5.0$ & $10\pm3$ & $9.12$ \\
\hline
\end{tabular}
\end{center}
\end{table}

Firstly, we show the peak counts from KiDS-450 in the upper panel of Fig.~1. The data are shown as points, their 
error bars have been calculated using a bootstrap sampling of individual KiDS-450 observation tiles, and the 
solid line is our best-fit theoretical model. The lower panel shows the residual between the data and this 
prediction. The corresponding peak numbers are also shown in Table~1. Secondly, Fig.~2 shows our 
fiducial constraints on $\Omega_{\rm m}$ and $\sigma_{\rm 8}$ in 
comparison with the results from the KiDS-450 cosmic shear tomographic 2PCF analysis (Hildebrandt et al. 2017). 
In addition, we show the pre-Planck CMB constraints (WMAP9+ACT+SPT, Calabrese et al. 2017), and the 
Planck CMB constraints ``TT+lowP'' (Planck Collaboration et al. 2016a).  

\begin{figure*}
\includegraphics[width=0.8\textwidth, angle=0]{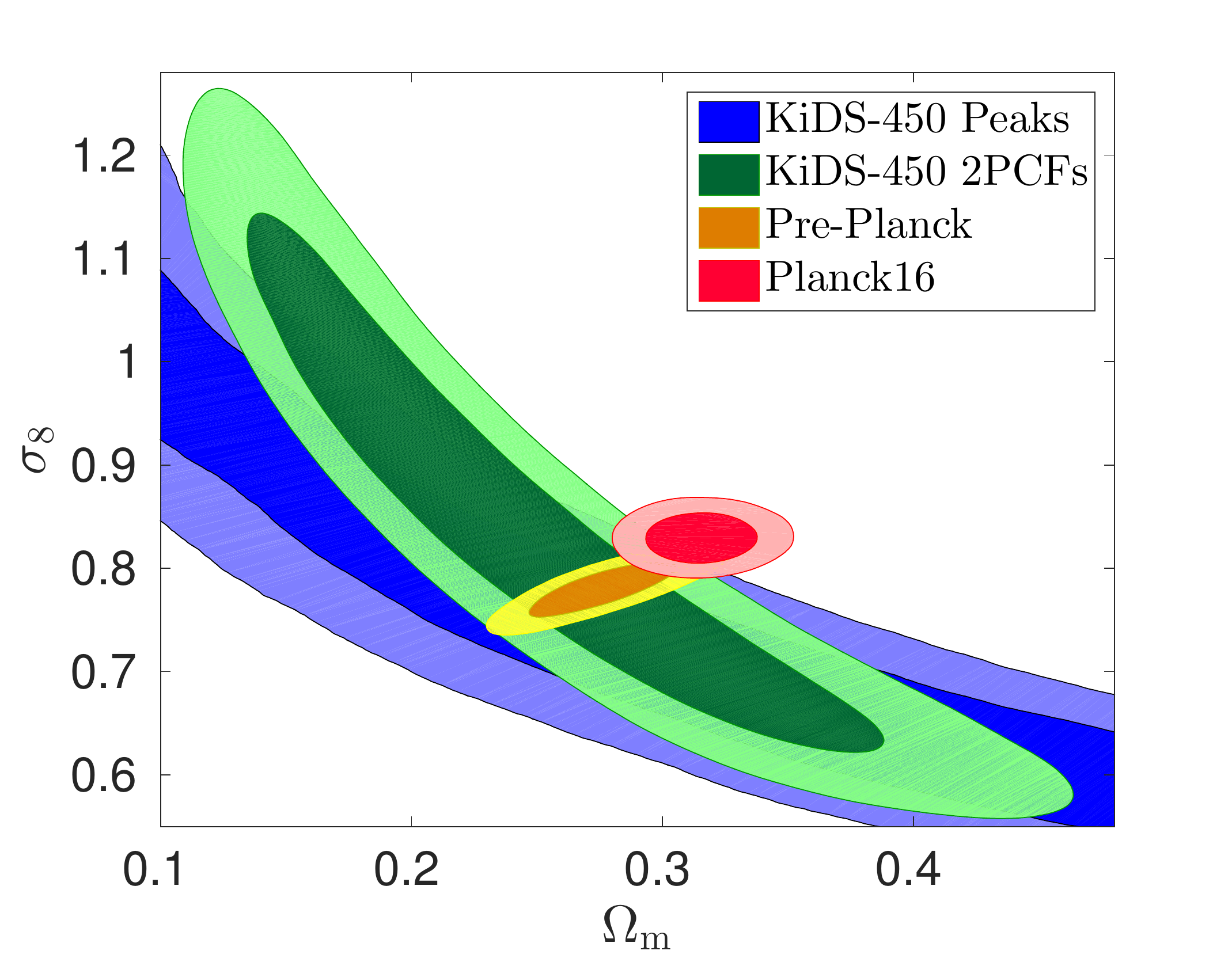}
\caption{The comparison for the constraints on $(\Omega_{\rm m}, \sigma_{\rm 8})$ between the fiducial
WL peak analysis (blue) and the results from the cosmic shear tomography from KiDS-450 (green). The constraints
from pre-Planck CMB measurement (yellow) and Planck 2016 (red) are also overplotted. The contours are $1\sigma$
and $2\sigma$ confidence levels, respectively.}
\label{fig:kids450_cosmology}
\end{figure*}

\begin{figure*}
\vspace*{-2.0cm}
\includegraphics[width=0.8\textwidth, angle=0]{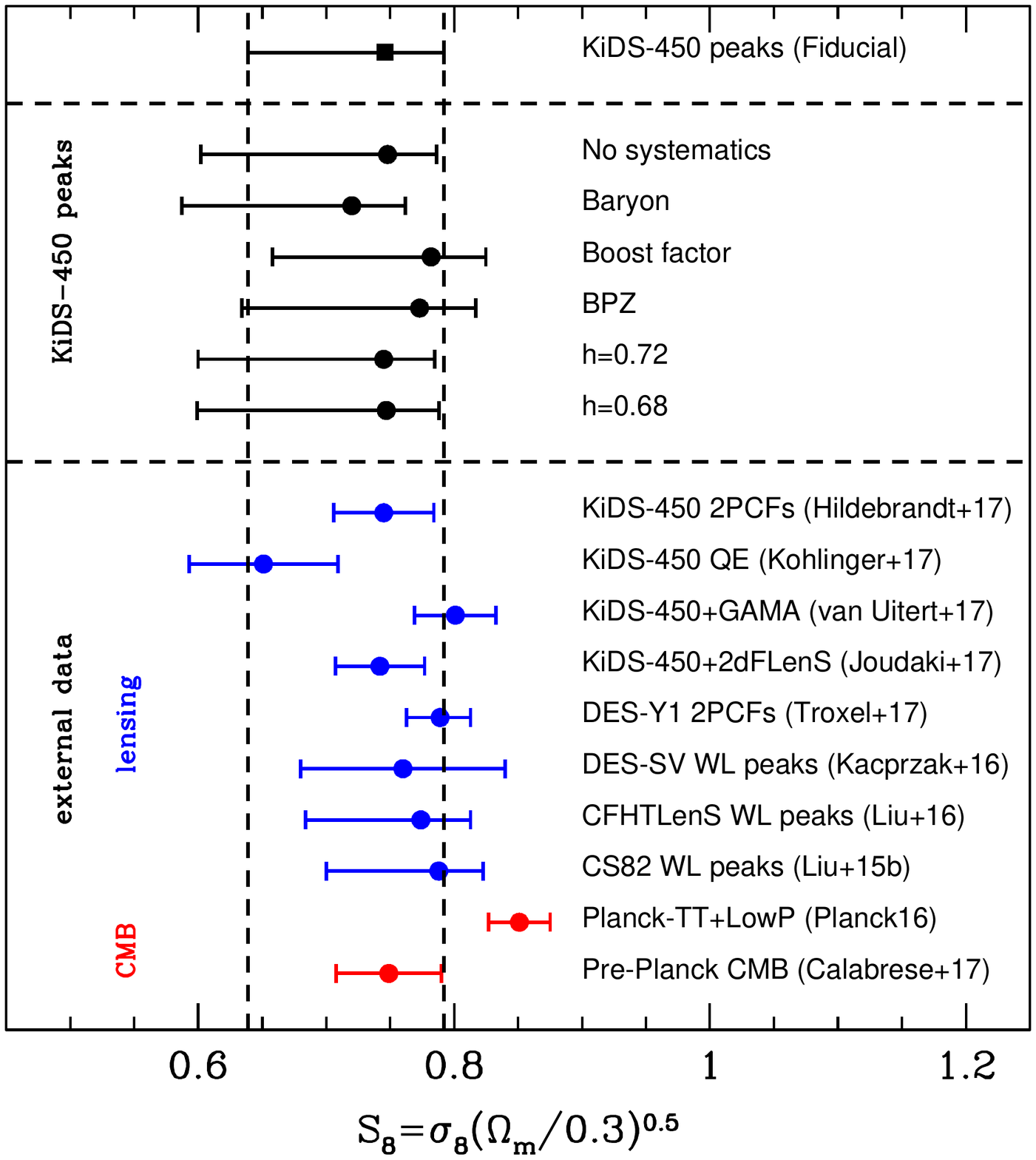}
\vspace*{-3.5cm}
\caption{Constraints on $S_{\rm 8}$ from our WL peak analysis, including various systematic tests, compared to 
various estimates from the literature measurements.}
\label{fig:kids450_S8}
\end{figure*}

From Fig.~1 and 2, we can see that the results from our WL peak analysis are an accurate representation 
of the KiDS-450 data, and that they are consistent with the cosmological constraints reported using a 2PCFs 
analysis of the same dataset. Both methods return constraints that agree well with pre-Planck CMB 
measurements. Furthermore it can be seen that, the degeneracy relation has 
a somewhat flatter slope than that from tomographic 2PCFs measurements. This difference means that our analysis 
has great potential to be used in a manner that is complementary to cosmic shear correlation analysis, as a 
joint analysis may provide tighter cosmological constraints than is possible with either analysis alone.

Finally, comparison with Planck CMB measurements reveals a tension similar to that reported in previous KiDS 
studies. This tension is quantified in the following section.

\subsection{Comparison of $S_{\rm 8}$ values}

Due to the strong degeneracy between $\Omega_{\rm m}$ and $\sigma_{\rm 8}$ from WL analyses, cosmological 
constraints are often characterised via the single quantity 
$\Sigma_8=\sigma_{\rm 8}(\Omega_{\rm m}/0.3)^{\alpha}$, where the 
index $\alpha$ is indicative of the slope of the degeneracy direction. When performing cosmic shear 2PCFs 
analyses this degeneracy is typically found to have a slope of $\alpha\sim 0.5$. 
As such, $\Sigma_8$ is frequently re-defined as $S_{\rm 8}=\sigma_{\rm 8}(\Omega_{\rm m}/0.3)^{0.5}$. 
In either case, with a freely varying or fixed value $\alpha$, this characterisation parameter can be 
constrained better than $\Omega_{\rm m}$ and $\sigma_{\rm 8}$ separately. 
Given the frequent use of $S_{\rm 8}$ rather than $\Sigma_8$ in the literature, we first calculate $S_8$ 
and subsequently calculate $\Sigma_{\rm 8}$, fitting for the free parameter $\alpha$. 

Using our fiducial WL peak analysis, we find $S_{\rm 8}=0.746^{+0.046}_{-0.107}$. This value is in agreement 
to that from cosmic shear tomographic 2PCFs analysis, which gives 
$S_{\rm 8}=0.745^{+0.039}_{-0.039}$ (Hildebrandt et al. 2017). To show the robustness of the results, we 
explore the impact (on our estimated $S_{\rm 8}$) of the various systematic effects which 
were accounted for in our model, and of some systematic effects external to our model. After these tests, we 
then also compare our $S_{\rm 8}$ estimates to additional constraints from the literature. 

\subsubsection{Testing systematic effects}

We first ignore all the measurement and astrophysical systematics, and estimate $S_{\rm 8}$ in the absence 
of our boost factor and baryonic effect corrections. This allows us to obtain 
a no-systematics estimate of $S_{\rm 8}=0.748^{+0.038}_{-0.146}$. This value is included in Fig.~3, 
and is indicative of how our estimate of $S_{\rm 8}$ changes under consideration of these two systematic effects. 
Interestingly, we can see that our fiducial measurement of $S_{\rm 8}$ is largely unchanged here. This is because 
of the compensation of the boost effect and the baryonic effect to be shown in the following. We note 
also that, for both of these estimates (and in fact for all our 
estimates of $S_{\rm 8}$), the error bars are strongly asymmetric. This is due primarily to the 
different degeneracy direction compared with the assumed slope of $\alpha = 0.5$. 
Indeed, fitting with a free $\alpha$ results in a much more symmetric uncertainty estimate (see Sect.~5.2). 
Moreover, the seemingly larger error bars in the case of no systematics is mainly due to the 
different degeneracy direction from $\alpha=0.5$. With the fitting $\alpha$, the probability distribution 
of $\Sigma_8$ is much more symmetric and the errors are indeed smaller in the no-systematics case than that 
of our fiducial analyses. 

\begin{figure*}
\includegraphics[width=0.8\textwidth]{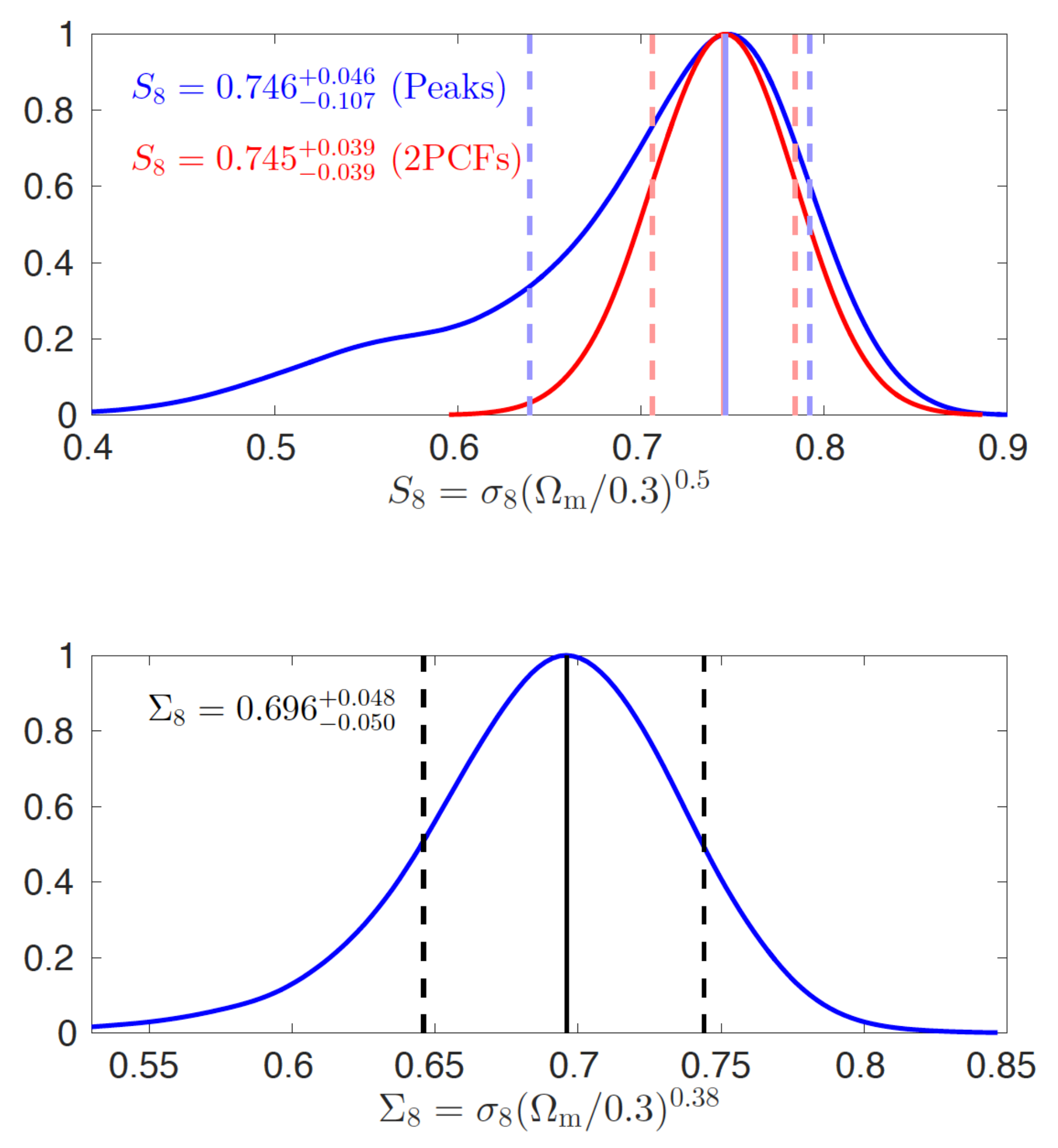}
\caption{Upper panel: The marginalised probability distribution of $S_{\rm 8}$ for KiDS-450 WL peak 
statistics (blue) and the cosmic shear tomographic 2PCFs analysis (red). Lower panel: The marginalised 
distribution of $\Sigma_{\rm 8}$ for KiDS-450 WL peak statistics.}
\label{fig:S8_distribution}
\end{figure*}

Considering only the boost effect, with the modified model described at length in Appendix~D, we find  
$S_{\rm 8}=0.782^{+0.043}_{-0.124}$. This shows that the boost factor pushes $S_{\rm 8}$ to higher 
values, and leads to a marginal reduction in uncertainty.  

Testing the influence of baryonic effects by freeing the $A$ parameter without including the boost 
effect, we find $S_{\rm 8}=0.720^{+0.042}_{-0.133}$. This is made by marginalising over $A$, and is 
also shown in Fig.~3. This estimate is $\sim3.8\%$ lower than the no-systematics value, and is marginally higher 
than might be expected from previous simulation studies (see, e.g., Osato et al. 2015). Nonetheless the effect 
is minor. However it is relevant to note that in the future this will not be the case. Future large WL surveys 
will provide sufficient area that WL peak counts will increase by order of magnitude. We expect that our 
self-calibration method will be particularly useful, allowing both a significant reduction in cosmological 
parameter constraint biases as well as valuable information about baryonic physics. 

The above analyses show that the two systematics move the $S_8$ estimate in opposite directions. As a 
result, when both are considered in our fiducial analyses, their effects are largely canceled out and 
the $S_8$ value is nearly unchanged comparing to the case of no systematics. 

We also assess the impact of redshift uncertainties. To do this, we carry out the peak analysis using the 
posterior redshift distribution P(z) returned by \textsc{BPZ}. Here we do not include the boost factor and 
baryonic effects, and our results are compared to our no-systematics estimate. This test returns a 
value $S_{\rm 8}=0.773^{+0.044}_{-0.139}$, and is shown in Fig.~3 as KiDS-450 peak (\textsc{BPZ}). 
This $S_{\rm 8}$ estimate is marginally higher than our no-systematics analysis, primarily because the mean 
of \textsc{BPZ} redshift distribution is lower than that of DIR. This is in agreement with the 
analysis of Hildebrandt et al. (2017), who observe a similar effect in cosmic shear constraints of $S_{\rm 8}$.

We also test how sensitive our estimate of $S_{\rm 8}$ is to the variation of the mean redshift of the 
bootstrapped DIR sample. We select the two bootstrap realisations with the most different mean 
estimates comparing to the one used in our main studies. Specifically, the difference in the mean redshift 
is $\Delta \langle z \rangle=+0.036$ and $\Delta \langle z \rangle=-0.037$, respectively. Correspondingly, 
the obtained values of $S_8$ are $S_{\rm 8}=0.744^{+0.039}_{-0.147}$ and $0.750^{+0.039}_{-0.136}$, 
respectively. The results are consistent with our no-systematics estimate within the statistical errors, 
indicating a negligible bias from the DIR photo-z uncertainties. 

Finally, in our analysis we have assumed a reduced Hubble constant $h=0.7$. However, recent results from 
Planck CMB temperature and polarization analyses suggest that $h$ may be smaller than our assumed value. 
To estimate the effect of a change in $h$ on our results, we perform two additional measurements of 
$S_{\rm 8}$ assuming $h=0.68$ and $h=0.72$. For the no-systematic cases, the derived parameters 
are $S_{\rm 8}=0.747^{+0.041}_{-0.148}$ and $S_{\rm 8}=0.745^{+0.040}_{-0.145}$, for 
$h=0.68$ and $h=0.72$ respectively. Again, these results are consistent with our fiducial estimate and 
indicate that our results are robust to modest variations in $h$. 

\subsubsection{External constraints}

When comparing our $S_{\rm 8}$ constraints with those from previous CMB temperature and polarization 
measurements, we find very good agreement with pre-Planck CMB-based constraints from Calabrese et al. (2017). 
However, similar to the tomographic 2PCFs analyses, our result is lower than the CMB
measurement from Planck ($S_{\rm 8}=0.851\pm0.024$, Planck Collaboration 2016a) at the level
of $\sim 2.0\sigma$. Fig.~3 shows these results and those from other KiDS-450 measurements, 
the Dark Energy Survey Year One (DES-Y1) cosmic shear measurement, and previous WL peak analyses, in 
comparison to our fiducial estimate and our various systematic tests from Sect.~5.1.1.

Our estimate of $S_{\rm 8}$ is consistent with all previous KiDS analyses, within $1\sigma$ uncertainties. 
To demonstrate this, we highlight the following results in particular.  
K\"ohlinger et al. (2017) use power spectrum analysis to estimate $S_{\rm 8}$, finding $S_{\rm 8}=0.651\pm0.058$. 
Combining cosmic shear measurements from KiDS-450 with galaxy-galaxy
lensing and angular clustering from GAMA, van Uitert et al. (2017) obtained $S_{\rm 8}=0.801\pm0.032$. In a
parallel analysis, Joudaki et al. (2017b) found $S_{\rm 8}=0.742\pm0.035$ using KiDS-450 cosmic shear
measurements with galaxy-galaxy lensing and redshift space distortion from the 2-degree Field Lensing Survey
(2dFLenS, Blake et al. 2016) and the Baryon Oscillation Spectroscopic Survey (BOSS, Dawson et al. 2013).

Moreover, our estimate of $S_{\rm 8}$ is also consistent with the recent results from DES-Y1. 
Troxel et al. (2017) report a cosmic shear based estimate of $S_{\rm 8}=0.789^{+0.024}_{-0.026}$, 
which is again in good agreement with the value presented here. 

We also compare our results to previous WL peak analyses in the literature, finding good agreement. 
Liu et al. (2015b) use CS82 data and find $S_{\rm 8}=0.788^{+0.035}_{-0.088}$. 
They also fit for a free $\alpha$, finding a lower value than $\alpha=0.5$ assumed by $S_{\rm 8}$.  
Liu et al. (2016) use CFHTLenS to constrain $f(R)$ theory using WL peak statistics. While they do not report 
$S_{\rm 8}$ directly, we are able to utilise their 
WL peak catalogue to estimate $S_{\rm 8}$ for their sample, finding $S_{\rm 8}=0.774^{+0.039}_{-0.090}$. 
Finally, Kacprzak et al. (2016) use DES-SV to study the abundance of shear peaks with $0<\nu<4$, identified 
in aperture mass maps. They constrain cosmological parameters using a suit of simulation templates with $158$ 
models with varying $(\Omega_{\rm m}, \sigma_{\rm 8})$ (Dietrich \& Hartlap 2010). They find 
$S_{\rm 8}=0.76\pm0.074$, with uncertainty derived by marginalising over the shear multiplicative bias and the 
error on the mean redshift of the galaxy sample. The constraints from these studies are marginally 
higher than our results, while being nonetheless consistent with our fiducial result within uncertainties. 

We conclude that our results are consistent with the pre-Planck CMB measurement of 
Calabrese et al. (2017), other KiDS-450 measurements, DES-Y1 cosmic shear and other WL peak analyses. 
The $\sim 2.0\sigma$ tension with Planck CMB measurements is again seen here. 

\subsection{Parameter degeneracy}\label{sec: param degen}

As shown in Fig.~2, our $(\Omega_{\rm m}, \sigma_{\rm 8})$ degeneracy direction is 
somewhat flatter than that present in 2PCFs analyses. This difference, we argue, results in significantly 
asymmetric uncertainties on our estimate of $S_{\rm 8}$. We demonstrate this clearly in the upper panel of 
Fig.~4, where we show the marginalised probability distribution of $S_{\rm 8}$ for our fiducial WL peak 
analysis (blue) and cosmic shear tomographic 2PCFs analysis (red). Our distribution is clearly heavily skewed, 
with a long tail toward the lower values of $S_{\rm 8}$. 

As this tail is clearly an artefact caused by the use of a fixed $\alpha=0.5$, we now explore how our estimates 
change when we fit with a freely varying $\alpha$; that is, we fit for $\Sigma_{\rm 8}$ rather than $S_{\rm 8}$. 
We derive the best-fit $\alpha\approx 0.38$ using the values of $(\Omega_{\rm m}, \sigma_{\rm 8})$ that 
are within $1\sigma$ confidence level of the constraints (the dark-blue region in Fig.~2). The smaller $\alpha$ 
reflects the flatter contours from our peak analyses than that from 2PCFs, consistent with the visual 
inspections. With the fitted $\alpha$, we then calculate the distribution of $\Sigma_{\rm 8}$ from the 
obtained constraints on $(\Omega_{\rm m}, \sigma_{\rm 8})$. The result is shown in lower panel of Fig.~4. 
It is seen that the distribution is significantly more symmetric than the distribution of $S_{\rm 8}$. With 
the best-fit $\alpha$, our final estimate is $\Sigma_{\rm 8}=0.696^{+0.048}_{-0.050}$.

We note that our constraint on $\alpha$ is similar to that recovered from cluster count analyses 
(Vikhlinin et al. 2009; Rozo et al. 2010; Planck Collaboration 2016b). These all find 
smaller $\alpha$ although they vary somewhat: Vikhlinin et al. (2009) find $\alpha\approx0.47$ 
from analyses of X-ray clusters; Rozo et al. (2010) find $\alpha\approx 0.41$ using MaxBCG analysis; and 
studies of SZ clusters find $\alpha\approx0.3$ (Planck Collaboration 2016b). The variations could be due 
to systematically different masses and redshifts probed by these different studies. It is interesting 
to note that the non-tomography high-SNR shear peak analyses of Dietrich \& Hartlap (2010) with simulation 
templates also obtain a flatter degeneracy direction. Each of these studies is broadly 
consistent with our best-fit $\alpha\sim 0.38$, which is expected due to the significant correlation 
between high-SNR WL peaks and massive clusters of galaxies. 

\section{Conclusions}\label{sec:conc}

We derive cosmological constraints from a WL peak count analysis using $450~\rm deg^2$ of KiDS data. As 
shape noise is the dominant source of uncertainties in our analysis we adopt the theoretical model of 
Fan et al. (2010), which takes into account the various effects of shape noise in modelling peak counts. 

We begin by testing the applicability of this model. Comparing its predictions with WL peak counts from 
simulations of different cosmologies (Appendix~A), we find good agreement between the model and our 
simulations. We also test the Gaussian approximation for the residual shape noise used in the 
model (Appendix~B), again finding consistent results. Finally, we perform a mock KiDS analysis using 
a suite of simulations to validate our full analysis pipeline (Appendix~C), finding that our pipeline 
recovers the input cosmology consistently.

After verifying both the model and our pipeline, we estimate our `fiducial' cosmological constraints using 
the DIR calibrated redshift distribution (Hildebrandt et al. 2017) and high-SNR peaks $(3<\nu<5)$, accounting 
for the influence of boost factor and baryonic effects. We explore other systematics, including projection 
effects of LSS, shear measurement bias, and photo-z errors, and conclude that these are insignificant for the 
WL peak analysis performed here. We explore the effect of intrinsic alignments (IA), finding it to have negligible
impact on shape noise variance and therefore on our results. However, considering the cluster member
contamination, we find that the peak signal measurements may be affected if member galaxies have intrinsic
alignments within clusters. The existence of such alignments is, however, still debated within the
literature; we opt not to include it in our analysis. Further study of the effects of IA are nonetheless
of interest, and we leave this for future work.

We summarise our primary conclusions as follows: 

(1) For a flat $\Lambda$CDM cosmology, our fiducial cosmological constraint on $(\Omega_{\rm m}, \sigma_{\rm 8})$ 
from WL peaks is $S_{\rm 8}=\sigma_{\rm 8}(\Omega_{\rm m}/0.3)^{0.5}=0.746^{+0.046}_{-0.107}$. This is 
consistent with previous estimates, within KiDS, from cosmic shear tomographic 2PCFs analysis and shear 
peak counts. Our estimate is also consistent with previous WL peak studies from CFHTLenS, CS82, and DES-SV. 
Finally, our result is consistent with pre-Planck CMB results, although we find a 
tension of $\sim 2.0\sigma$ with the Planck CMB.

(2) We perform a quantitative analysis of a range of systematic effects, including photo-z errors and 
uncertainty in the Hubble constant $h$, finding that these are insignificant compared to 
the statistical uncertainties on our value of  $S_{\rm 8}$.

(3) We fit for the degeneracy slope of $(\Omega_{\rm m}, \sigma_{\rm 8})$ from our high-SNR peak studies, 
characterised by the index $\alpha$, finding a slope somewhat flatter than that found using cosmic shear 
2PCFs analysis. This raises the potential for WL peak analysis to be used alongside a 2PCFs analysis, thereby 
breaking part of the $(\Omega_{\rm m}, \sigma_{\rm 8})$ degeneracy. Fitting for our cosmological constraint 
with $\alpha$ as a free parameter, we find 
$\Sigma_{\rm 8}=\sigma_{\rm 8}(\Omega_{\rm m}/0.3)^{\alpha}=0.696^{+0.048}_{-0.050}$, with the 
best-fit $\alpha=0.38$.

Previous estimates of $\alpha$ using low- and medium-SNR shear peaks (Kacprzak et al. 2016) find a degeneracy 
direction similar to that of cosmic shear 2PCFs measurements. We argue that the primary complementarity 
with 2PCFs studies, therefore, lies in studying high-SNR peaks. However, as the number of high-SNR peaks is 
still relatively low, even in our $450~\rm deg^2$ sample, the statistical uncertainties 
remain considerably larger than those of low-SNR peaks. Future WL surveys, 
such as {\it Euclid} (Laureijs et al. 2011), LSST (Abell et al. 2009) and 
the Wide Field Infrared Survey Telescope (WFIRST\footnote{\url{http://wfirst.gsfc.nasa.gov/}}), will 
provide considerably larger samples of high-SNR peaks, and thus allow us to extract much more cosmological 
information from studies of this nature. However, achieving higher accuracy will come at a 
cost: much tighter control on systematic effects will be paramount. 

\section*{Acknowledgements}

We are thankful for the referee's encouraging comments and suggestions. The analyses are based 
on data products from observations made with ESO Telescopes at the La Silla Paranal Observatory under
programme IDs 177.A-3016, 177.A-3017 and 177.A-3018, and on data products produced by Target/OmegaCEN,
INAF-OACN, INAF-OAPD and the KiDS production team, on behalf of the KiDS consortium. OmegaCEN and the KiDS
production team acknowledge support by NOVA and NWO-M grants. Members of INAF-OAPD and INAF-OACN also
acknowledge the support from the Department of Physics \& Astronomy of the University of Padova, and of the
Department of Physics of Univ. Federico II (Naples). We thank Antony Lewis for the \textsc{CosmoMC} packages.
HYS acknowledges support from TR33 project ``The Dark Universe'' funded by the DFG. HHi is suported
by an Emmy Noether grant (No. Hi 1495/2-1) of the DFG. XKL, CZP and ZHF are supported in part by the NSFC of 
China under grants 11333001 and 11173001 and by Strategic Priority Research Program The Emergence of 
Cosmological Structures of the Chinese Academy of Sciences, grant No. XDB09000000. XKL also acknowledges 
the support from General Financial Grant from China Postdoctoral Science Foundation with Grant No. 2016M591006.
HHo acknowledges support from Vici grant 639.043.512, financed by the Netherlands Organisation for Scientific 
Research (NWO). JHD acknowledges support from the European Commission under a Marie-Sk{l}odwoska-Curie 
European Fellowship (EU project 656869). CH acknowledges support from the European Research Council under 
grant number 647112. KK acknowledges support by the Alexander von Humboldt Foundation. JM has received funding 
from the European Union's FP7 and Horizon 2020 research and innovation programmes under Marie Sklodowska-Curie 
grant agreement numbers 627288 and 664931. QW acknowledges the support from NSFC with Grant No. 11403035. 
Part of the N-body simulations are performed on the Shuguang cluster at Shanghai Normal 
University, Shanghai, China.

\clearpage

\appendix

\section{Cosmological dependence of WL peak model}

\begin{figure}
\vspace*{-2.0cm}\includegraphics[width=0.5\textwidth]{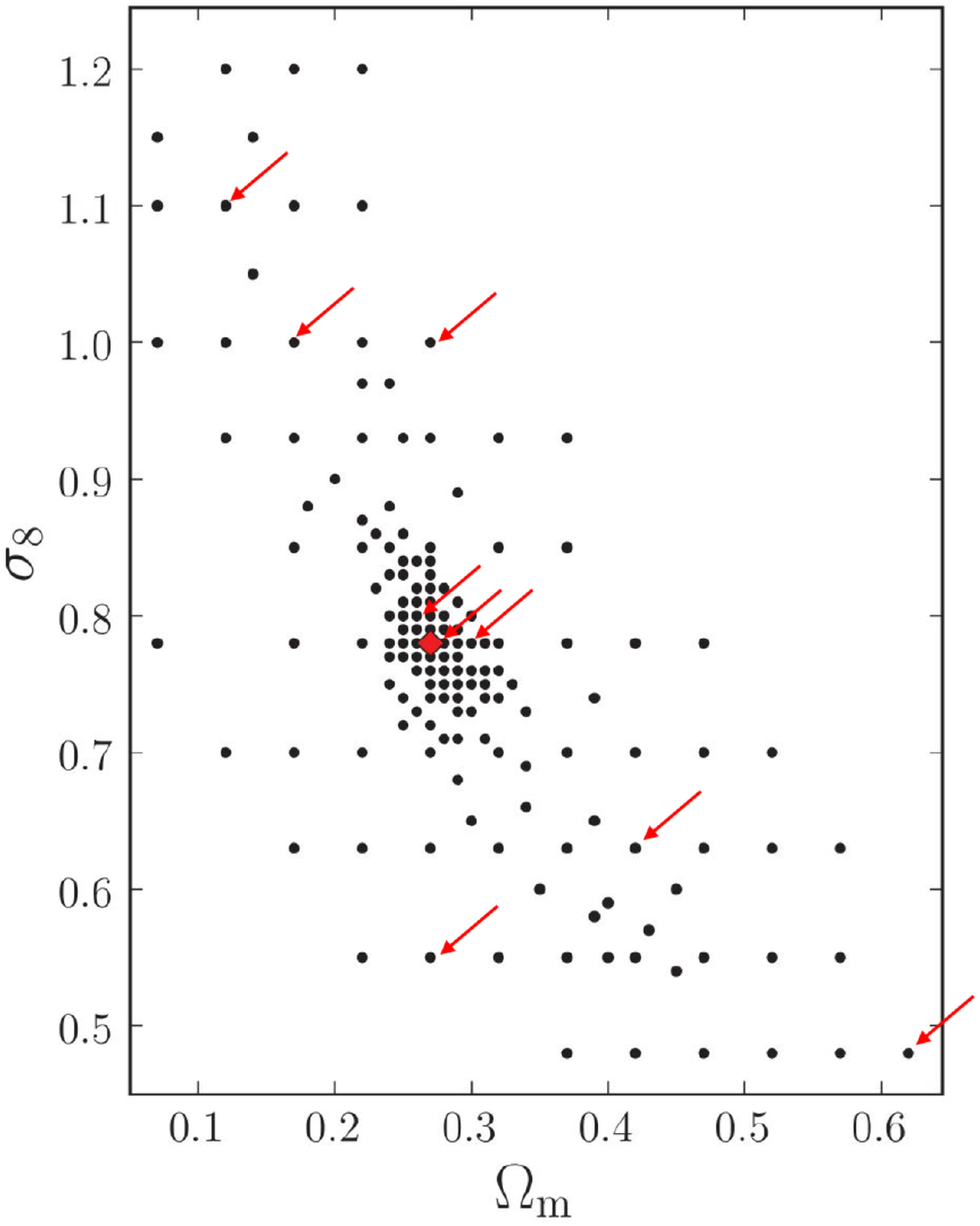}
\vspace*{-2.5cm}\caption{The spatial distribution of $158$ different cosmologies in the
$\Omega_{\rm m}-\sigma_{\rm 8}$ plane. The red diamond marks the fiducial cosmology at
$(\Omega_{\rm m}, \sigma_{\rm 8})=(0.27, 0.78).$ The red arrows denote the
cosmologies used for WL peak model tests.}
\label{fig:sim_select}
\end{figure}

\begin{figure*}
\includegraphics[width=0.8\textwidth]{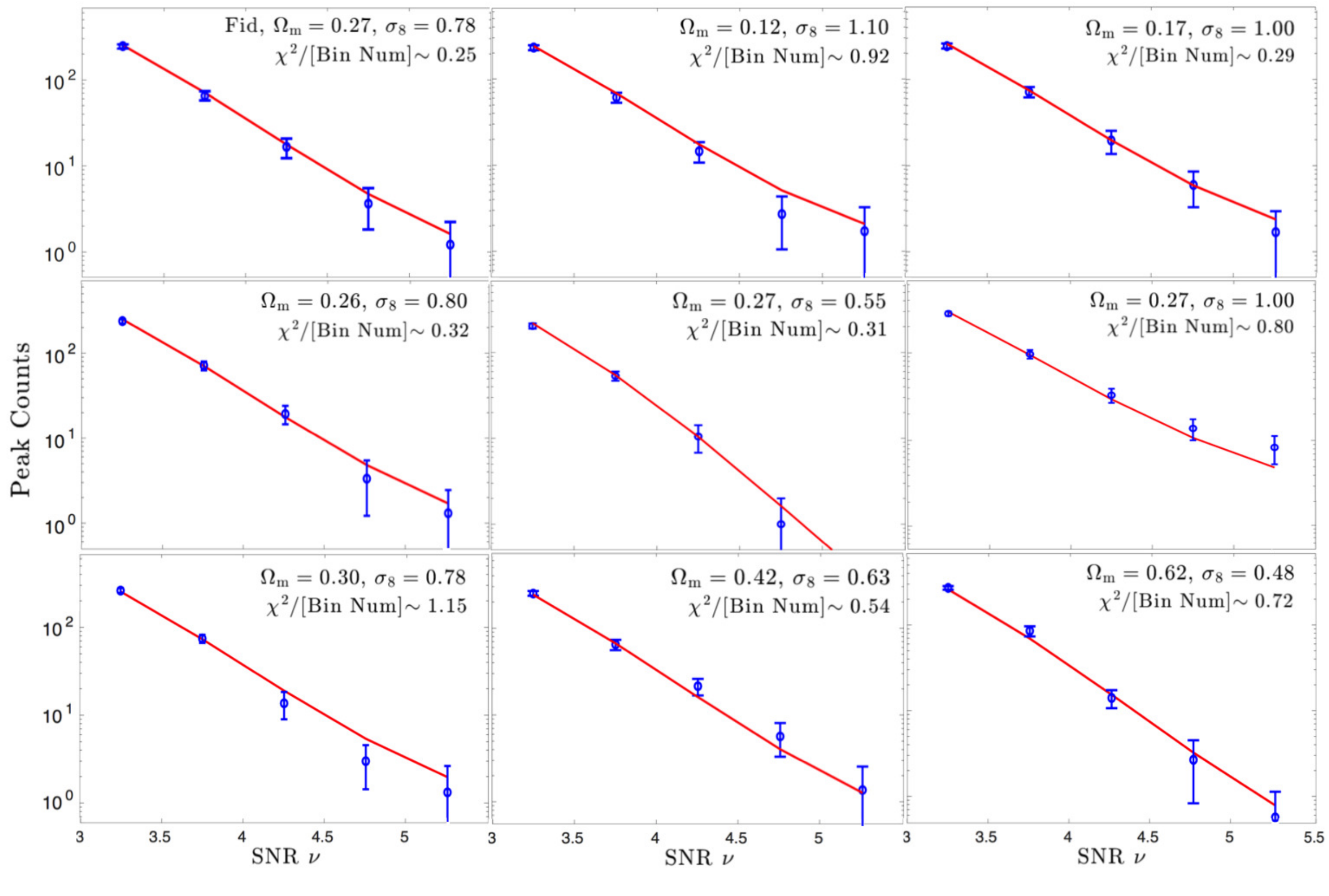}
\caption{The peak count distribution of the selected simulations. The corresponding solid lines are the
theoretical prediction with the input cosmological parameters.}
\label{fig:sim_peak}
\end{figure*}

Here we present tests of the Fan et al. (2010) peak model against simulations. From a suite of simulations 
with $158$ different cosmological models from Dietrich \& Hartlap (2010), each with a 
different $(\Omega_{\rm m}, \sigma_{\rm 8})$, we choose $9$ cosmological models (Fig.~A1) to perform our tests. 
We also include the `fiducial' $(\Omega_{\rm m}, \sigma_{\rm 8}) = (0.27,0.78)$ cosmological 
model from Dietrich \& Hartlap (2010), because of its increased sampling. 

Each of these simulations is stored in the form of single galaxy catalogue (containing position, redshift, and 
shear) where galaxies have been sampled uniformly over a $6\times6\rm~deg^2$ patch with number 
density $n_{\rm g}\sim25\rm~arcmin^{-2}$. From each catalogue, we generate a mock sample by randomly 
sampling galaxies to reproduce the galaxy number density and redshift distribution of our data set. 

For each non-fiducial cosmology, Dietrich \& Hartlap (2010) produce a single simulation box with $5$ 
different (random) lines of sight, with total area of $\sim180~{\rm deg}^2$. We then sample our 
galaxies $3$ times for each line of sight, thus generating $3$ sets of mocks for each model. After 
excluding boundaries the final on-sky area for each of our $8$ non-fiducial cosmologies 
is $\sim150~{\rm deg}^2$, each sample 3 times. For the fiducial cosmology, however, there are $35$ individual 
simulation boxes, each with $5$ different random LOS. For this cosmology, we 
sample our galaxy only once per line of sight, thus generating $35$ individual mock catalogues with a final 
on-sky area of $\sim1050~{\rm deg}^2$. We therefore end up with $295$ individual mock catalogues to analyse, 
generated from $215$ individual lines of sight across $43$ individual simulation boxes with $9$ different 
cosmologies. 

For each mock catalogue we perform a mass reconstruction and peak identification, and then fit the peak 
distribution with the theoretical predictions of Fan et al. (2010). Fig.~A2 shows the results of these 
fits for each of our $9$ cosmologies. In the figure, the symbols show the peak counts averaged:  
\begin{itemize}
  \item over $35$ mocks for the fiducial mode, and 
  \item over 3 mocks for the others.  
\end{itemize}
The uncertainties on the data points are the expected analytic uncertainties in peak counts given a 
survey area of $150~{\rm deg}^2$. We also present the value of $\chi^2/n_{\rm bins}$ for each model 
fit in the upper right of each panel.

In all cases we see that the model predications agree well with the simulation results. Note, however, that 
the simulation mocks here are somewhat idealised; for instance, there is no masking in these mocks. 
Nonetheless, they are sophisticated enough for the purpose of testing the peak model performance. 
In Appendix~C, we show analyses of mock images that replicate KiDS more accurately. 

\section{Residual noise properties}

\begin{figure*}
\includegraphics[width=0.6\textwidth]{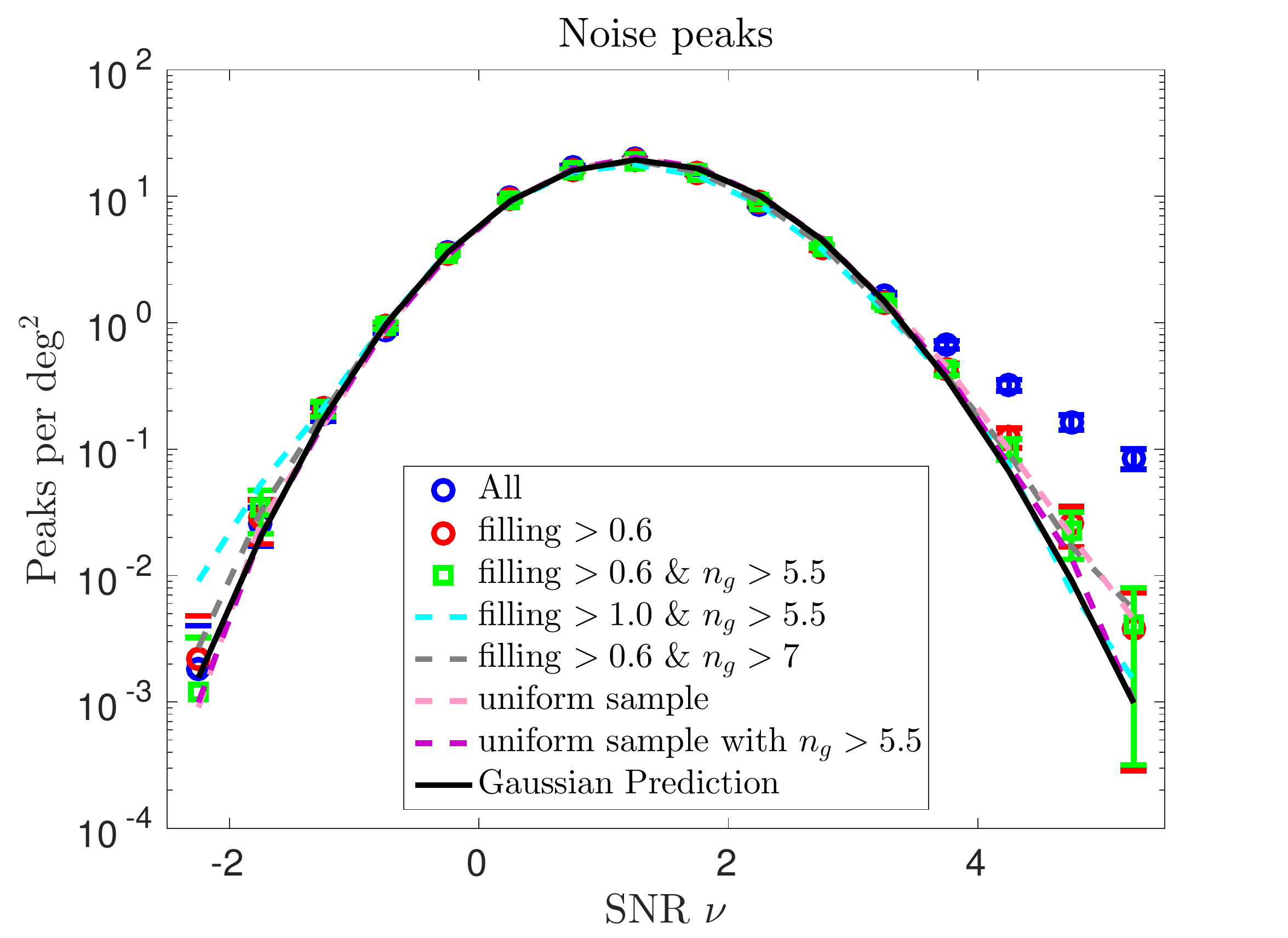}
\caption{Noise properties for different $n_{\rm g}$ and filling factors. 
The solid black line is the corresponding Gaussian prediction for the peak distribution. Blue symbols are
the results from noise maps without any selection. The other symbols and lines represent the results 
with different selection criteria. The error bars are evaluated from a bootstrap analysis.}
\label{fig:sim_noise}
\end{figure*}

One assumption within the peak model of Fan et al. (2010) is that the residual shape noise field is well 
described by a Gaussian random field. Van Waerbeke (2000) demonstrate that, when the effective number of 
source galaxies within the smoothing kernel is larger than $\sim10$, the residual shape noise is 
approximately Gaussian to a good degree. For the KiDS-450 dataset used here, 
$n_{\rm eff}\sim 7.5~{\rm arcmin}^{-2}$. Therefore, for a smoothing scale $\theta_G=2~{\rm arcmin}$ we 
expect that the Gaussian approximation for the noise field should be valid. However the source galaxy 
distribution varies from tile to tile, and within a tile the galaxy distribution is also truly random; there 
are far fewer source galaxies in regions that are heavily masked. We therefore need to set appropriate 
selection criteria to ensure the validity of the Gaussian noise approximation that we have assumed. 

From the noise maps described in Sect.~3.3, we analyse the one-point probability 
distribution function of the noise, and corresponding noise peak distribution, with different $n_{\rm g}$ 
and filling factor selection criteria. In Fig.~B1, the solid black line shows the shape of the assumed Gaussian 
peak noise distribution. Blue symbols are the results from noise maps without applying any selection criteria, 
and clearly shows some non-Gaussianity at high SNR. The other symbols and lines represent the results 
with different selection criteria as shown in the legend. Uncertainties on the data are estimated using 
a bootstrap analysis. From the figure we can see that, while the Gaussian approximation cannot describe the 
noise peaks well in the raw-counts case, by applying some modest selection criteria the approximation holds 
quite well. With a requirement of the filling factor $f>0.6$, which is designed to exclude the mask effects, 
the peak distribution is much closer to the Gaussian case. With $n_{\rm eff}>5.5~{\rm arcmin}^{-2}$, 
the results can be improved further. Applying even more stringent cuts, we are able to make the noise distribution 
converge on the Gaussian case almost perfectly, however this also causes a significant reduction in the number 
statistics. Finally we note that, if the galaxy distribution is a purely random selection of galaxies on-sky, 
the noise peak distribution becomes almost a perfect Gaussian when applying the 
simple $n_{\rm eff}>5.5~{\rm arcmin}^{-2}$ selection. In the realistic case where the galaxy 
distribution is unlikely to be a perfectly random sampling on-sky, the agreement with the assumed Gaussian 
distribution is acceptable (within uncertainties) when applying our modest selection criteria. Thus in our 
analysis, to ensure the validity of the Gaussian approximation while maintaining appropriate number statistics, 
we invoke two selection criteria on filling factor $f>0.6$ and effective number 
density $n_{\rm eff}>5.5~{\rm arcmin}^{-2}$.

\section{Mock analysis}

\begin{figure*}
\includegraphics[width=0.45\textwidth]{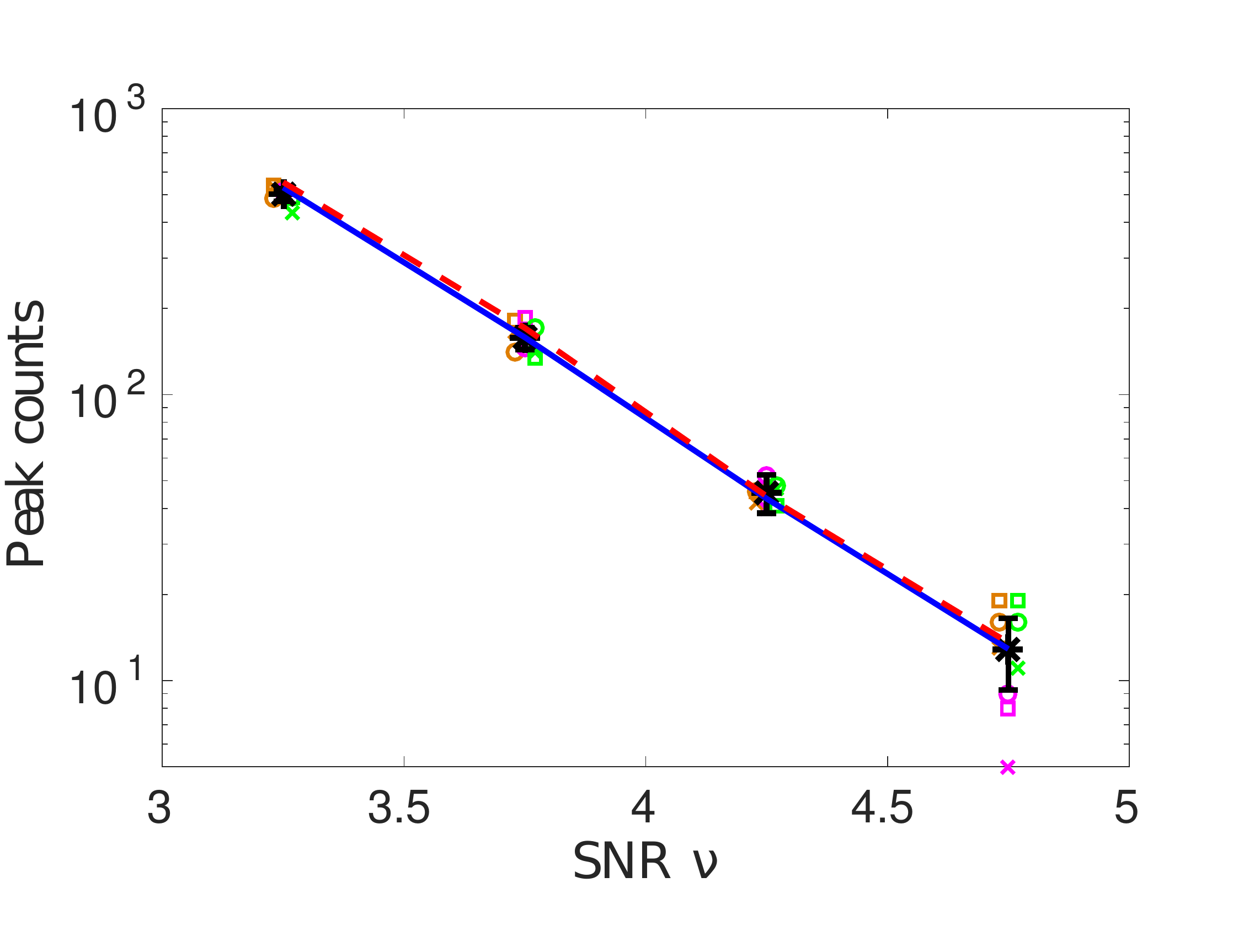}
\includegraphics[width=0.45\textwidth]{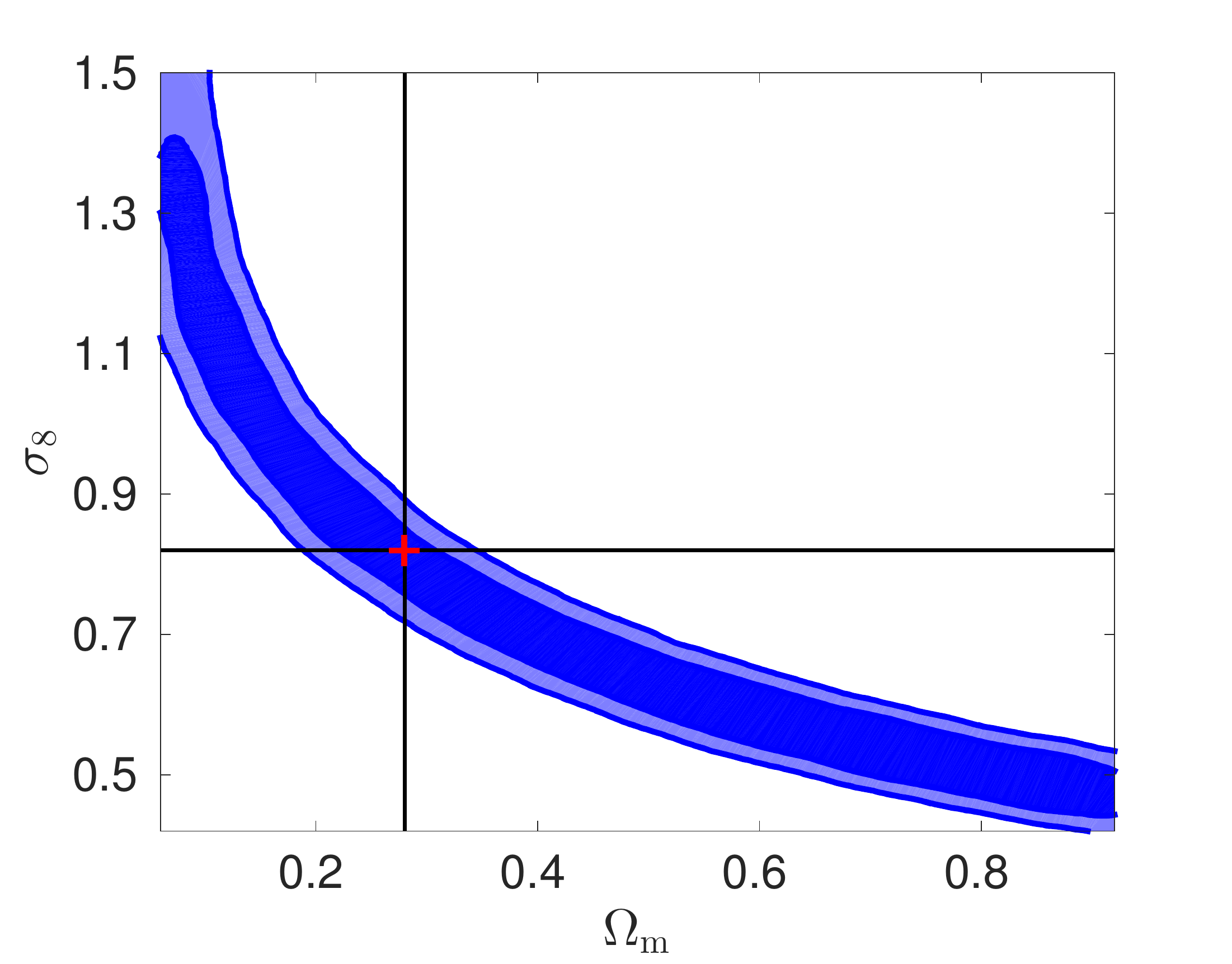}
\caption{The results of the mock analysis. Left panel: The peak count distribution of the KiDS-450 mock
data. The black `*' denotes the average value of the $3$ independent mocks. The error bars are the square
root of the diagonal terms of the covariance matrix. The nine sets of symbols with different colours correspond
to the $3$ independent mocks with $3$ different noise realizations. The solid blue line is the theoretical
prediction with the input cosmological parameters. The dashed red line is the peak count of the
smoothed convergence field including a uniform random noise field with the mean noise level of $9$ mocks.
Right panel: Cosmological constraints on $(\Omega_{\rm m}, \sigma_{\rm 8})$ derived from the KiDS-450 mock 
peak count. The red `+' denotes the input $(\Omega_{\rm m}, \sigma_{\rm 8})$.}
\label{fig:mock_peak}
\end{figure*}

Here we present our validation of the full analysis pipeline using mock KiDS-450 data constructed from our 
simulations. These mocks are generated from the ray-tracing simulations described in Liu et al. (2015b). 
Briefly, we run a large suite of N-body simulations and pad them together to redshift $z=3$ for ray-tracing 
calculations. Cosmological parameters in this simulation are chosen to be 
$(\Omega_{\rm m}, \Omega_{\Lambda}, \Omega_{\rm b}, h, \sigma_{\rm 8}, n_{\rm s})
=(0.28, 0.72, 0.046, 0.7, 0.82, 0.96)$. Each box is only used once, and so no repetitive structures occur; 
shifts and rotations of boxes are therefore not needed. From the simulations, we generate $96$ lensing 
maps each with an area of $3.5\times 3.5~{\rm deg}^2$, for a total area of $ 1176~\rm deg^2$. This 
allows us to create $3$ independent $449.7~\rm deg^2$ KiDS-like mocks\footnote{These mocks 
are different from the SLICS mocks used in previous KiDS publications (Harnois-D\'{e}raps \& Van Waerbeke 2015)}. 
For each mock we generate catalogues using $3$ different random rotations of galaxy intrinsic ellipticities, 
to produce three sets of shape noise, thus producing $9$ sets of mock catalogs to be used in this validation test.

The mock catalogue contains the position, observed ellipticity, weight, and redshift of each source therein.  
Each of these parameters is defined such that the mock is an appropriate representation of KiDS: 
\begin{itemize}
  \item The position and the shear measurement weight of each galaxy are taken to be the same 
as that of the KiDS-450 data, and so we are able to appropriately reproduce the KiDS masking in our mocks.
  \item Galaxy redshifts within the mock are generated by assigning a random value from the DIR redshift 
distribution of KiDS-450. 
  \item The observed ellipticity of each galaxy is constructed by combining the reduced shear and the 
intrinsic ellipticity. 
  \item The galaxy reduced shear $g$ is calculated by interpolating the lensing signals from the 
grids of the simulated lensing maps to the galaxy position (the interpolation is also done in the redshift 
dimension). 
  \item The intrinsic ellipticity is generated by keeping the amplitude of the observed ellipticity of the 
galaxy, but with its orientation being randomised.
\end{itemize} 
We then parse these mock catalogues through the same pipeline as we do the observed KiDS data to 
construct convergence maps, tile by tile, and produce the mock WL peak catalog. These individual 
catalogues are then used to derive cosmological constraints, as per KiDS. 

In Fig.~C1, the left panel shows the peak number distributions from the mock data. The symbols in the 
figure are coloured according to which of the $3$ independent simulated maps from which they originated. 
The $3$ symbol types within each colour are the results from different noise realizations. The 
black `*' denotes the average value per bin from the $9$ mocks. We estimate the covariance matrix by 
generating $10^4$ bootstrap samples by resampling the $9\times454$ tiles from the mocks. The error bars 
associated with each black `*' are derived using the RMS of the diagonal elements of the covariance matrix. 
The solid blue line is the theoretical prediction from Fan et al. (2010), with the input cosmological 
parameters. The dashed red line is the peak count of the smoothed convergence field from simulation 
including a uniform random noise field with the noise level of $9$ mocks, which is in good agreement with 
the theoretical prediction within $1\sigma$.

The right panel of Fig.~C1 shows the derived constraints on $(\Omega_{\rm m}, \sigma_{\rm 8})$ using the 
average peak counts from the mocks. The contours are $1\sigma$ and $2\sigma$ confidence levels, 
respectively. The red `+' denotes the input cosmological parameters, which are recovered excellently 
by the pipeline. We therefore conclude that the pipeline is performing well even when confronted with 
the complexity of real data. 

\section{Boost factor}

\begin{figure*}
\includegraphics[width=0.4\textwidth]{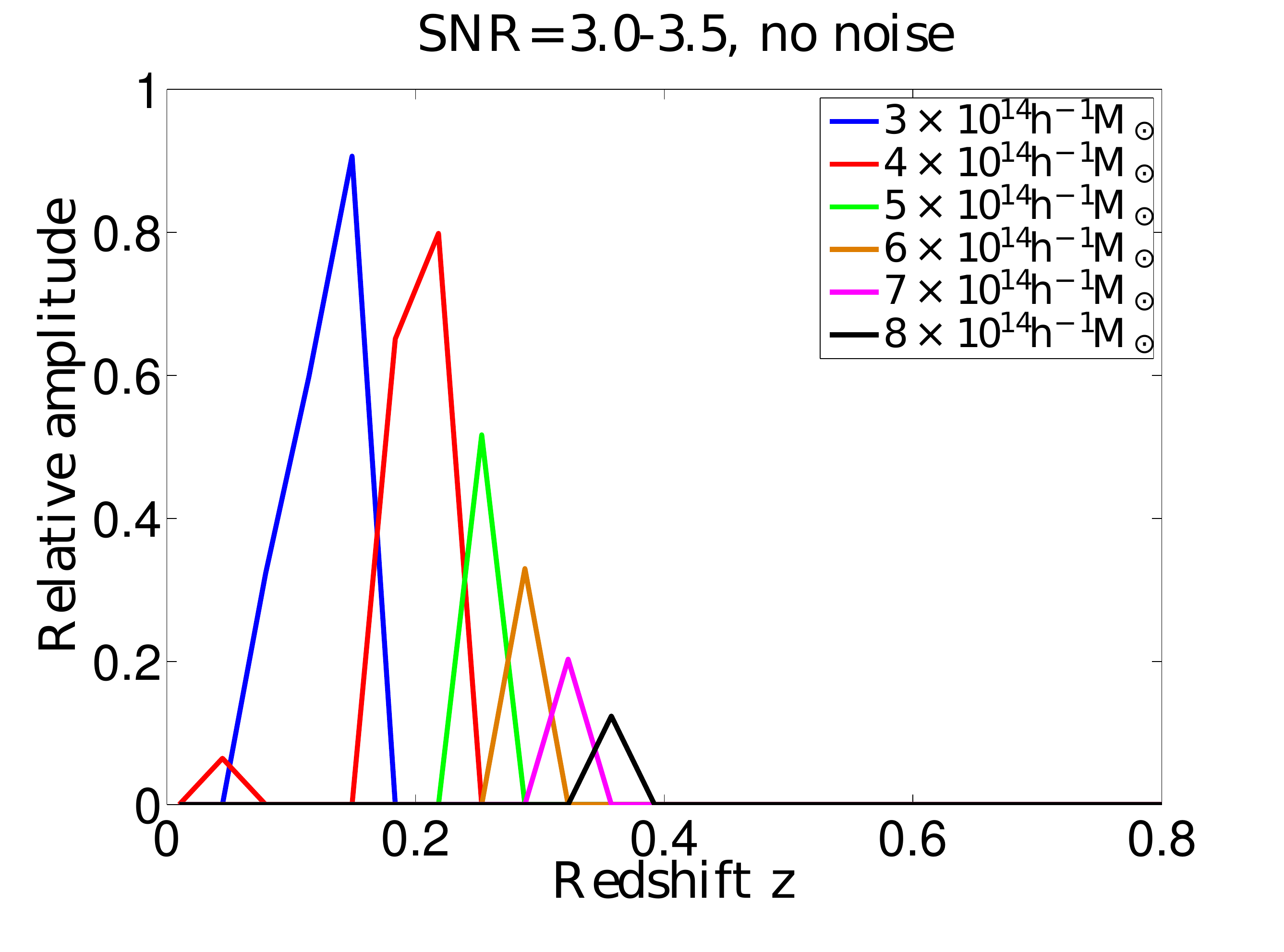}
\includegraphics[width=0.4\textwidth]{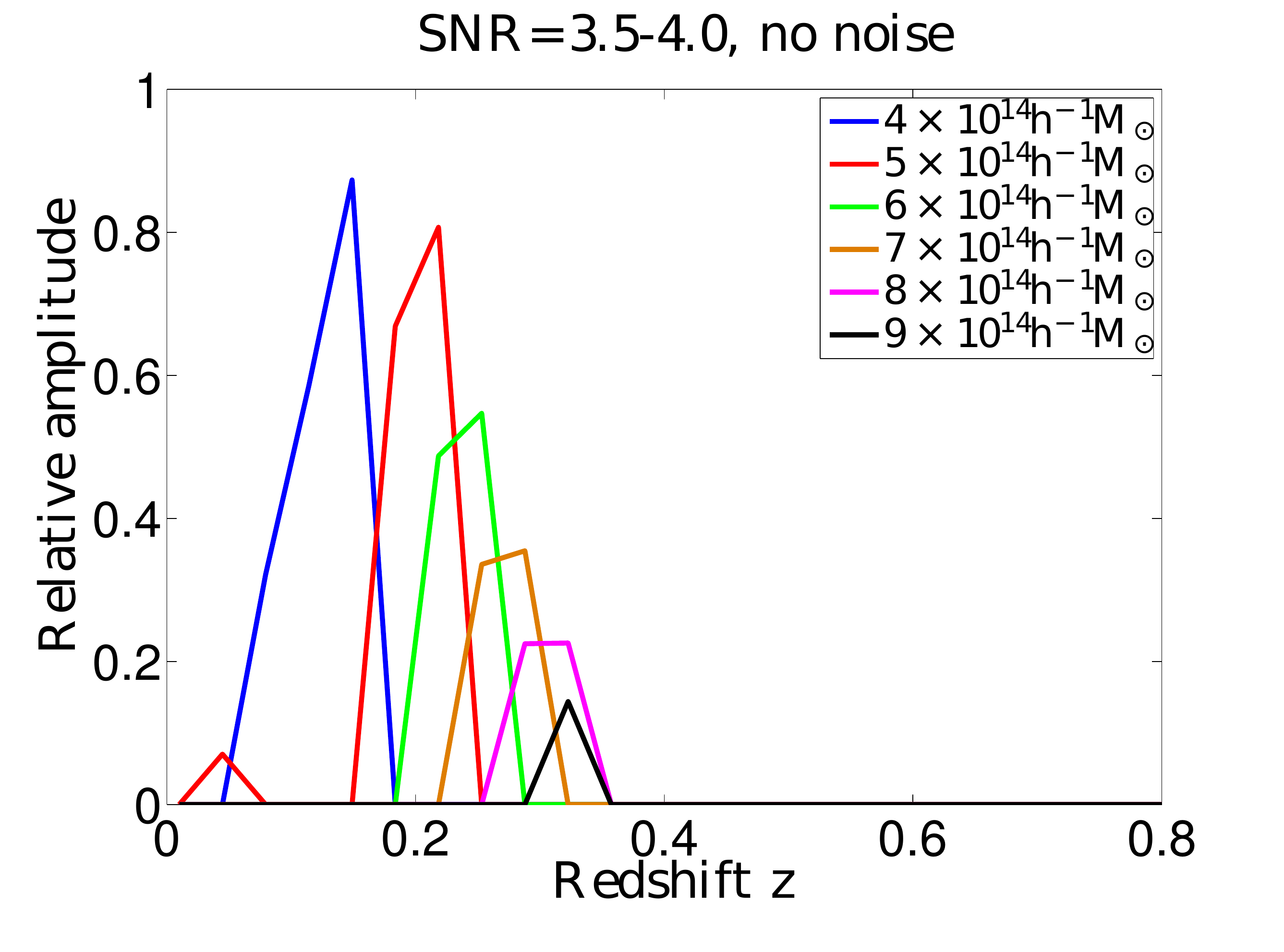}
\includegraphics[width=0.4\textwidth]{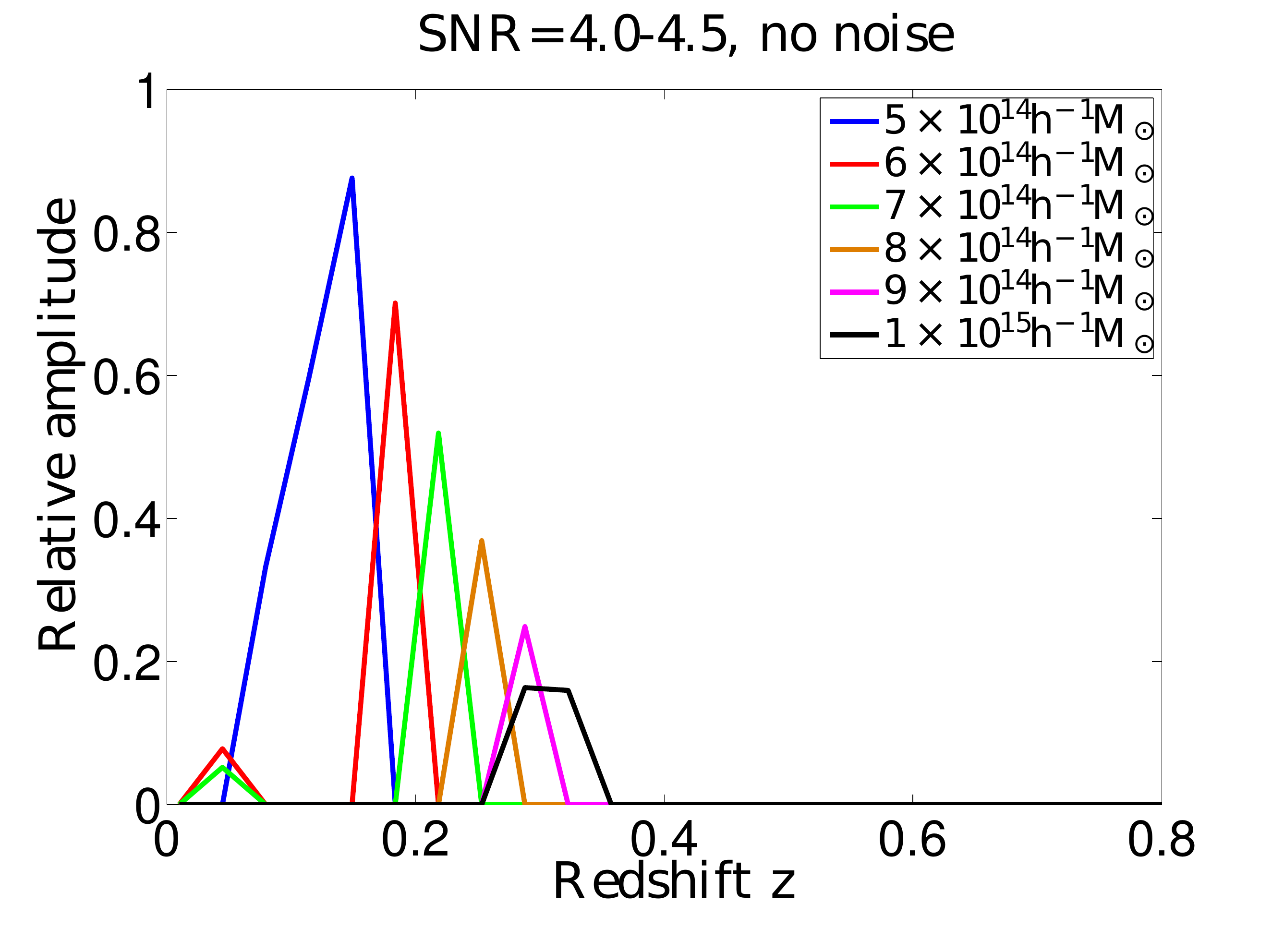}
\includegraphics[width=0.4\textwidth]{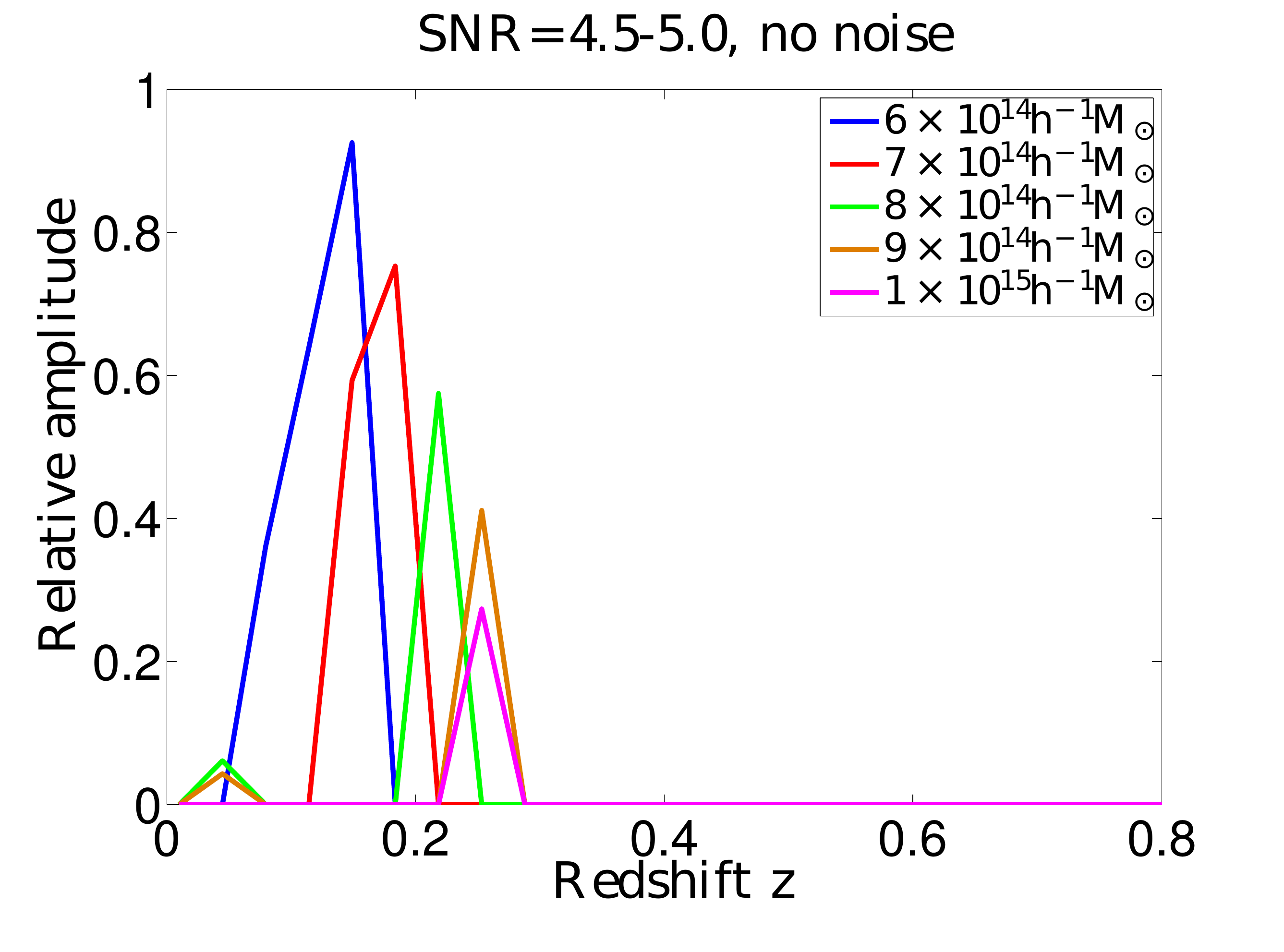}
\caption{The fractional contributions of clusters with different mass and redshift to WL peaks of different 
heights using the KiDS-450 source redshift distribution without the shape noise.}
\label{fig:noiseless}
\end{figure*}

\begin{figure*}
\includegraphics[width=0.4\textwidth]{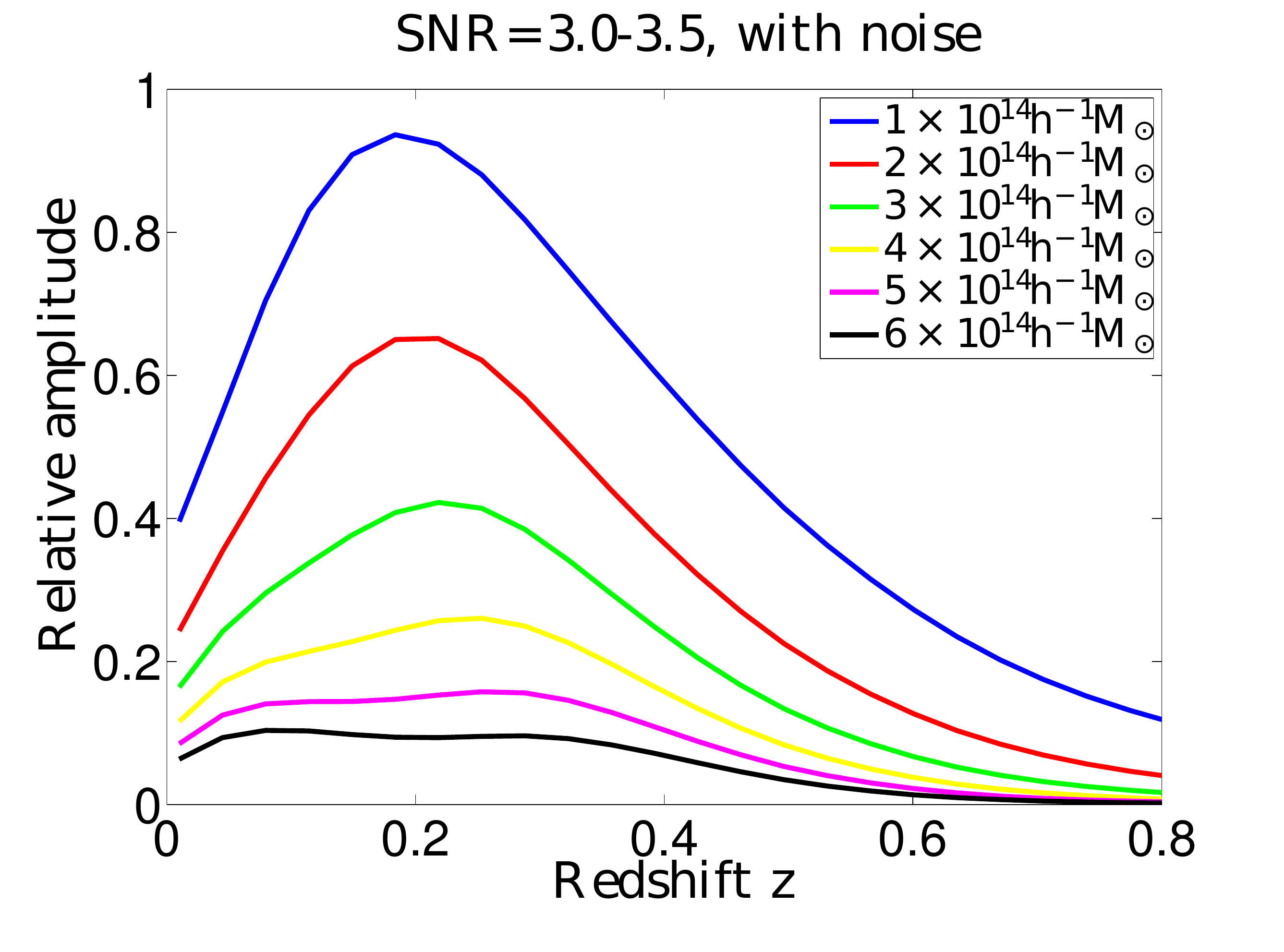}
\includegraphics[width=0.4\textwidth]{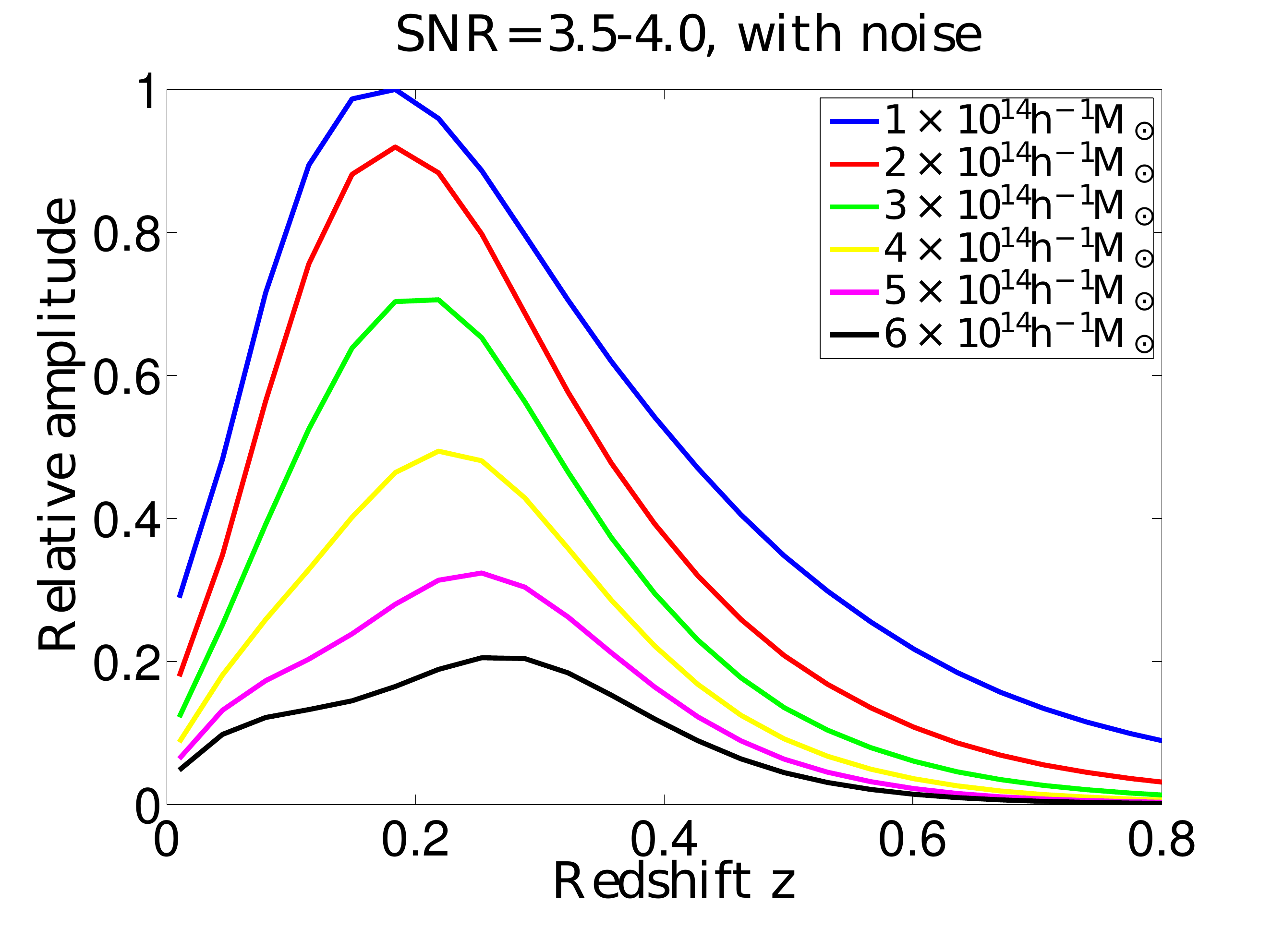}
\includegraphics[width=0.4\textwidth]{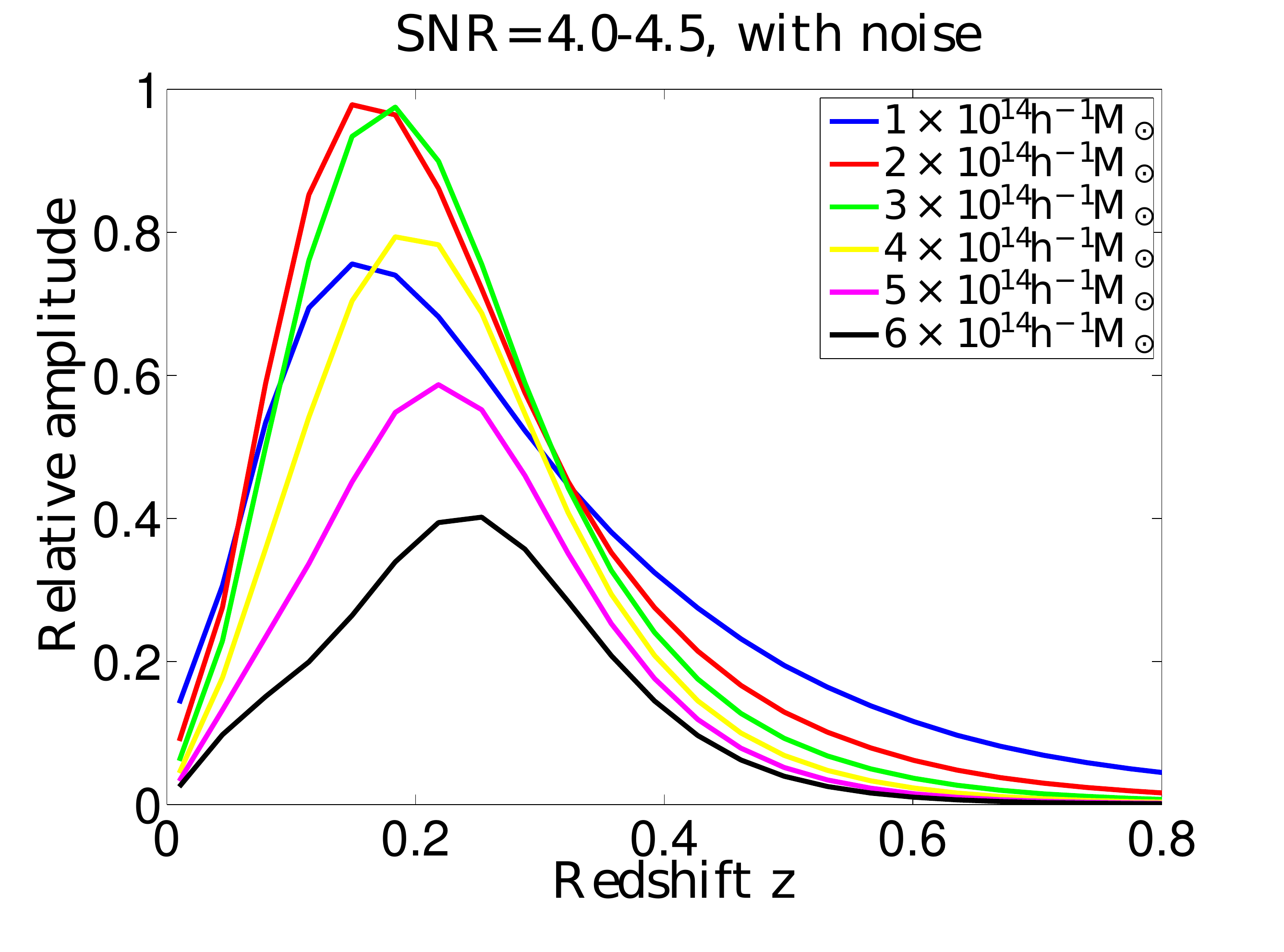}
\includegraphics[width=0.4\textwidth]{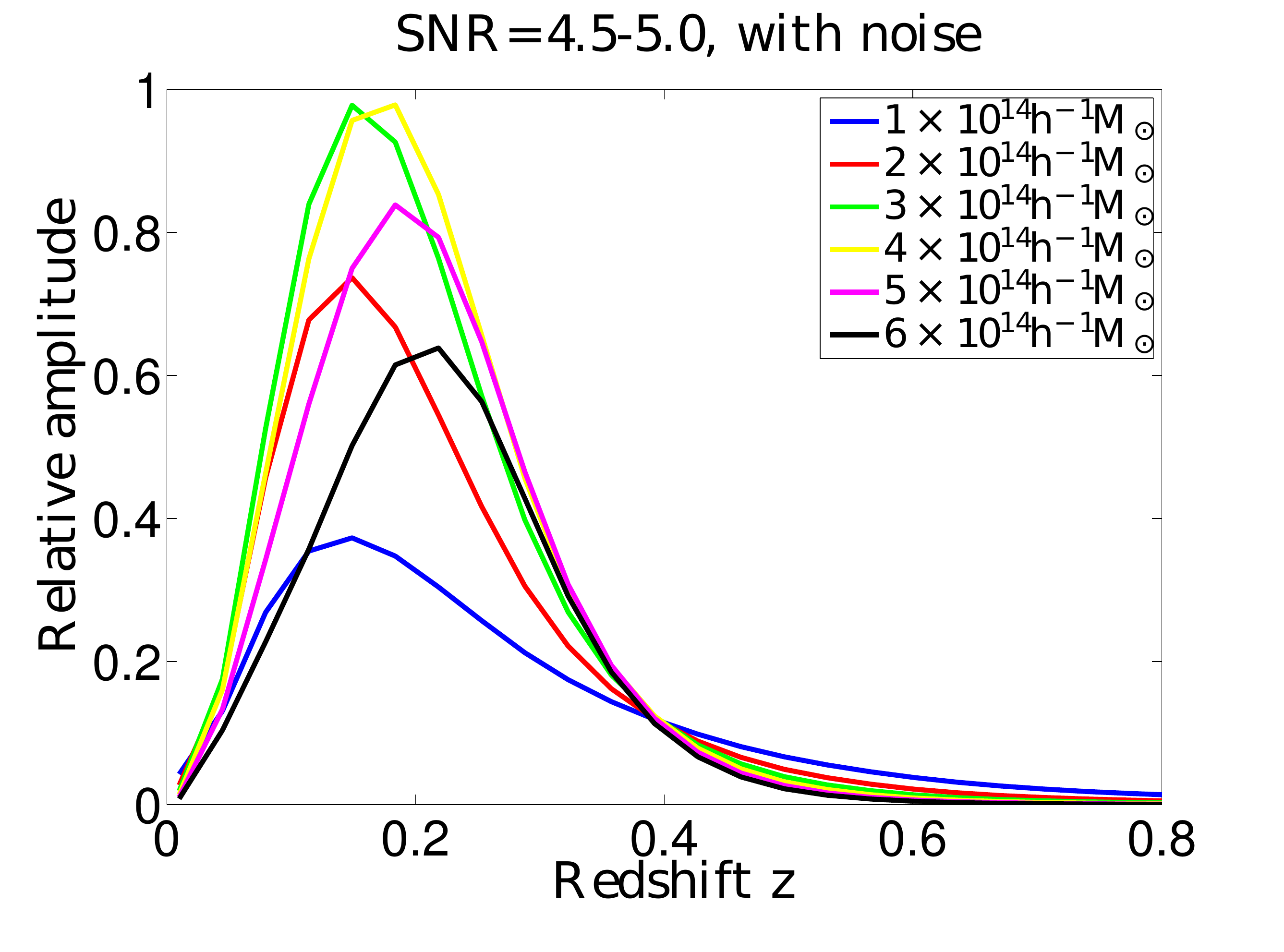}
\caption{The fractional contributions of clusters with different mass and redshift to WL peaks of different 
heights using the KiDS-450 source redshift distribution with KiDS-450-like shape noise.}
\label{fig:noise}
\end{figure*}

\begin{figure*}
\includegraphics[width=0.80\textwidth]{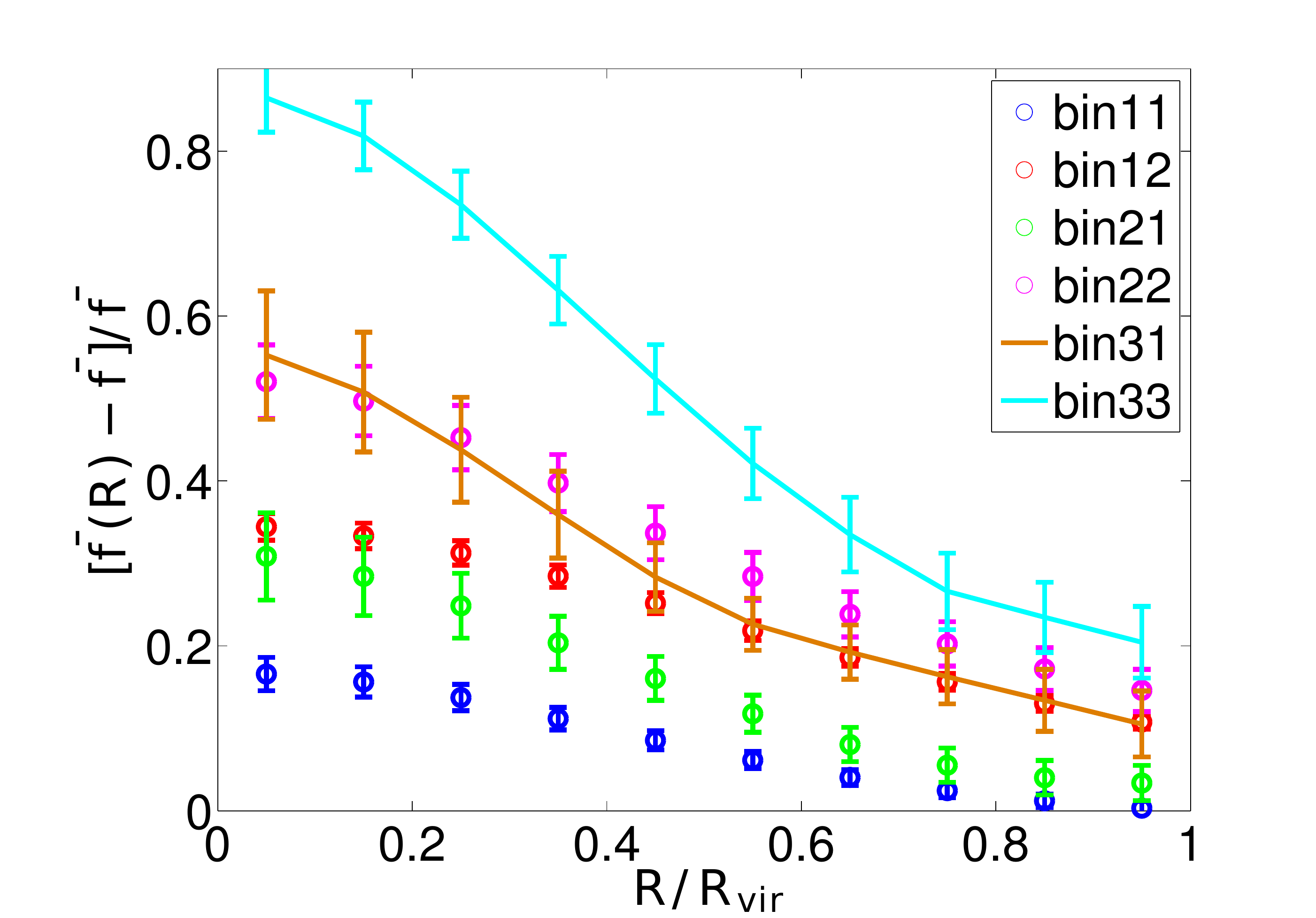}
\caption{Excess galaxy number density (filling factor) distribution around cluster candidates for each bin. 
The error bars on the mean are estimated from a bootstrap analysis.}
\label{fig:boost}
\end{figure*}

\begin{figure*}
\includegraphics[width=0.33\textwidth]{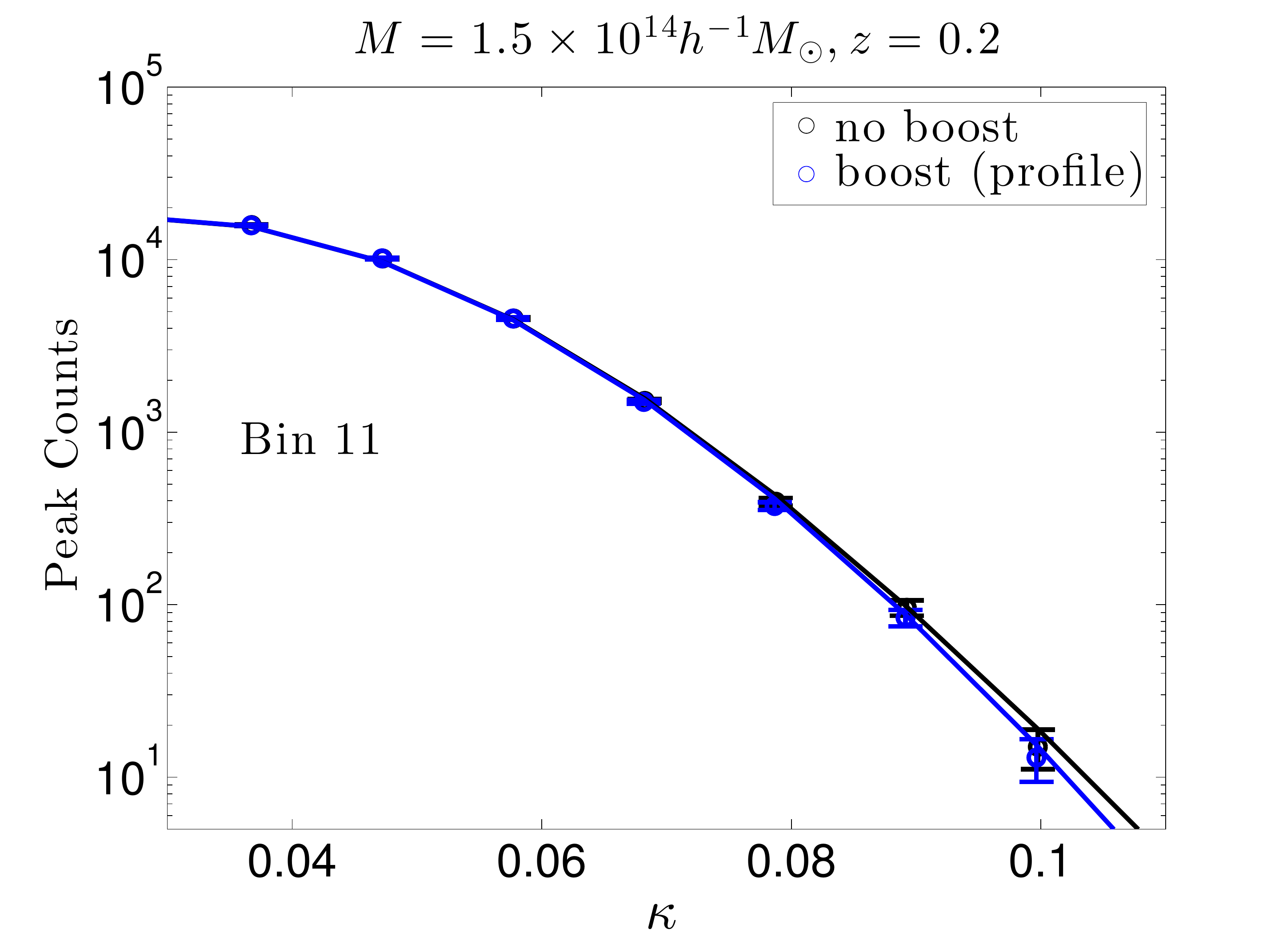}
\includegraphics[width=0.33\textwidth]{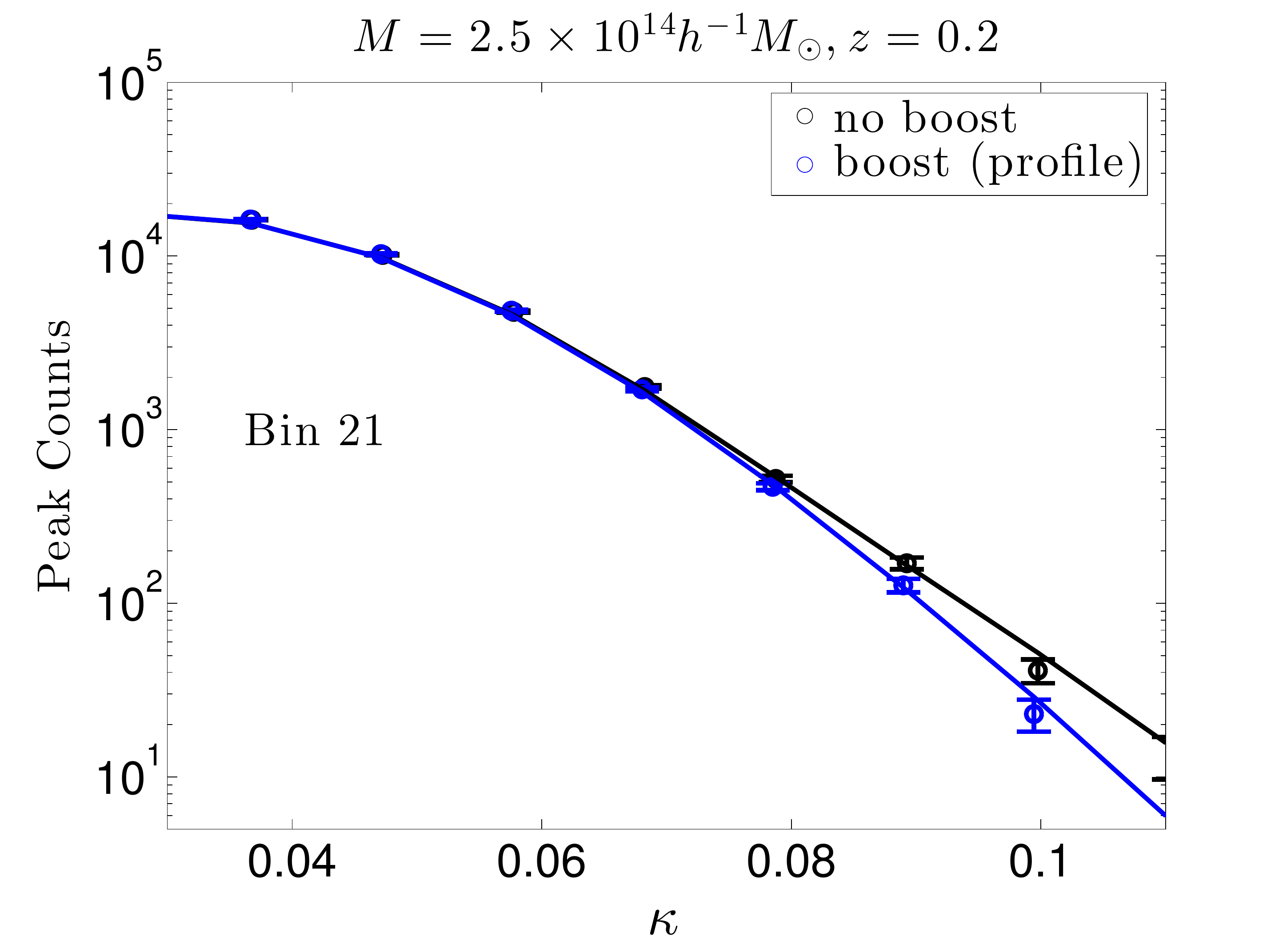}
\includegraphics[width=0.33\textwidth]{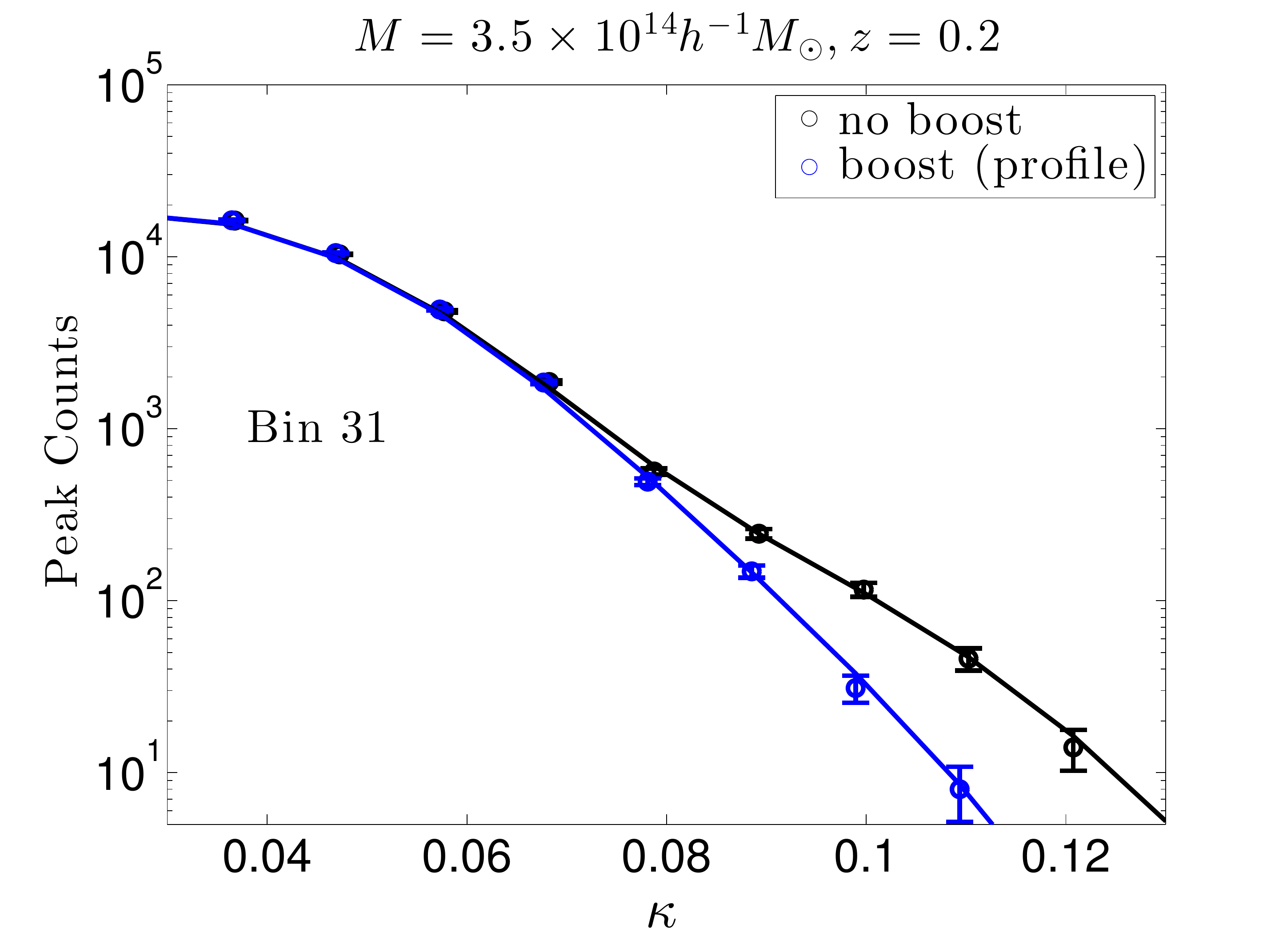}
\includegraphics[width=0.33\textwidth]{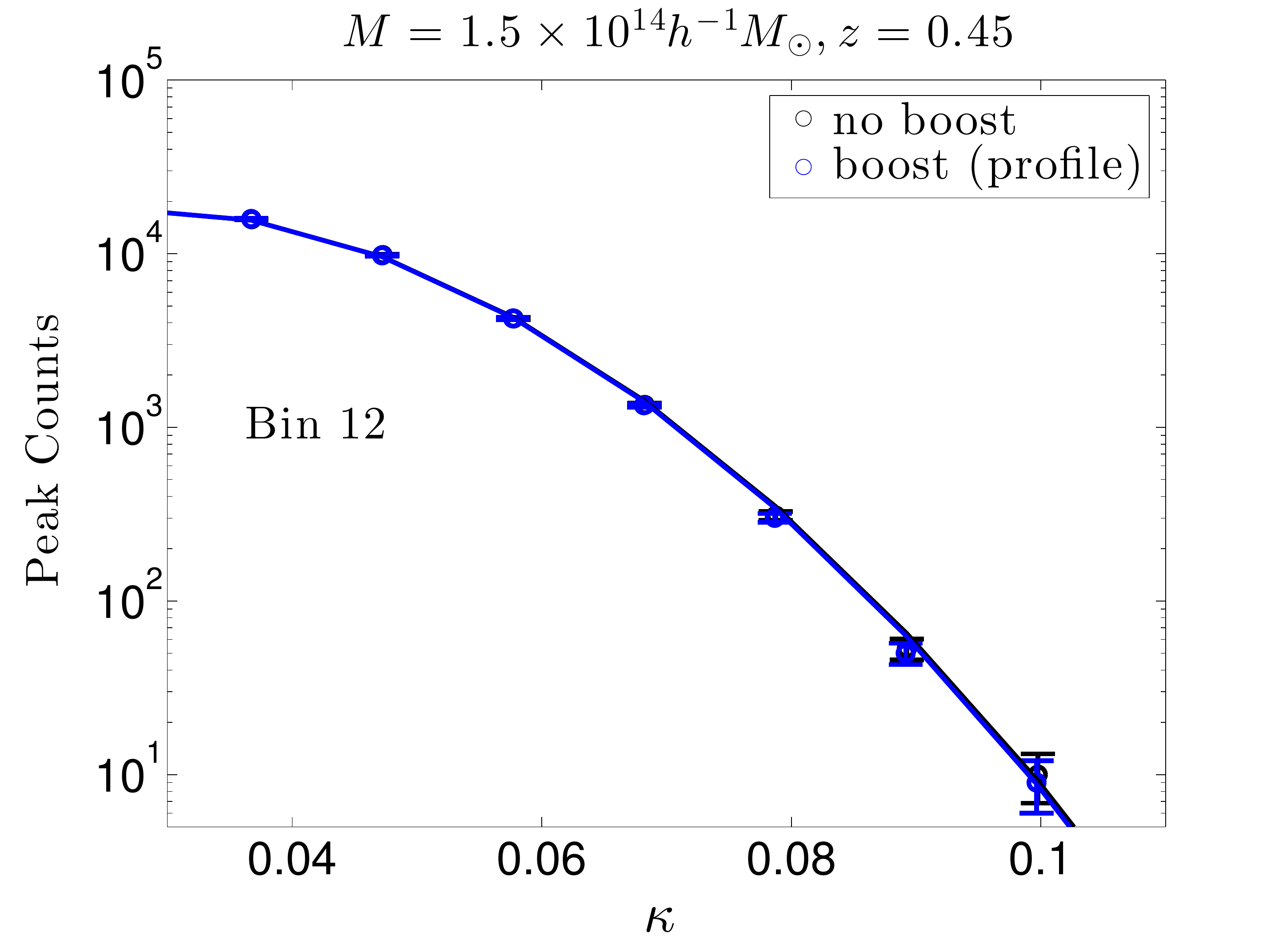}
\includegraphics[width=0.33\textwidth]{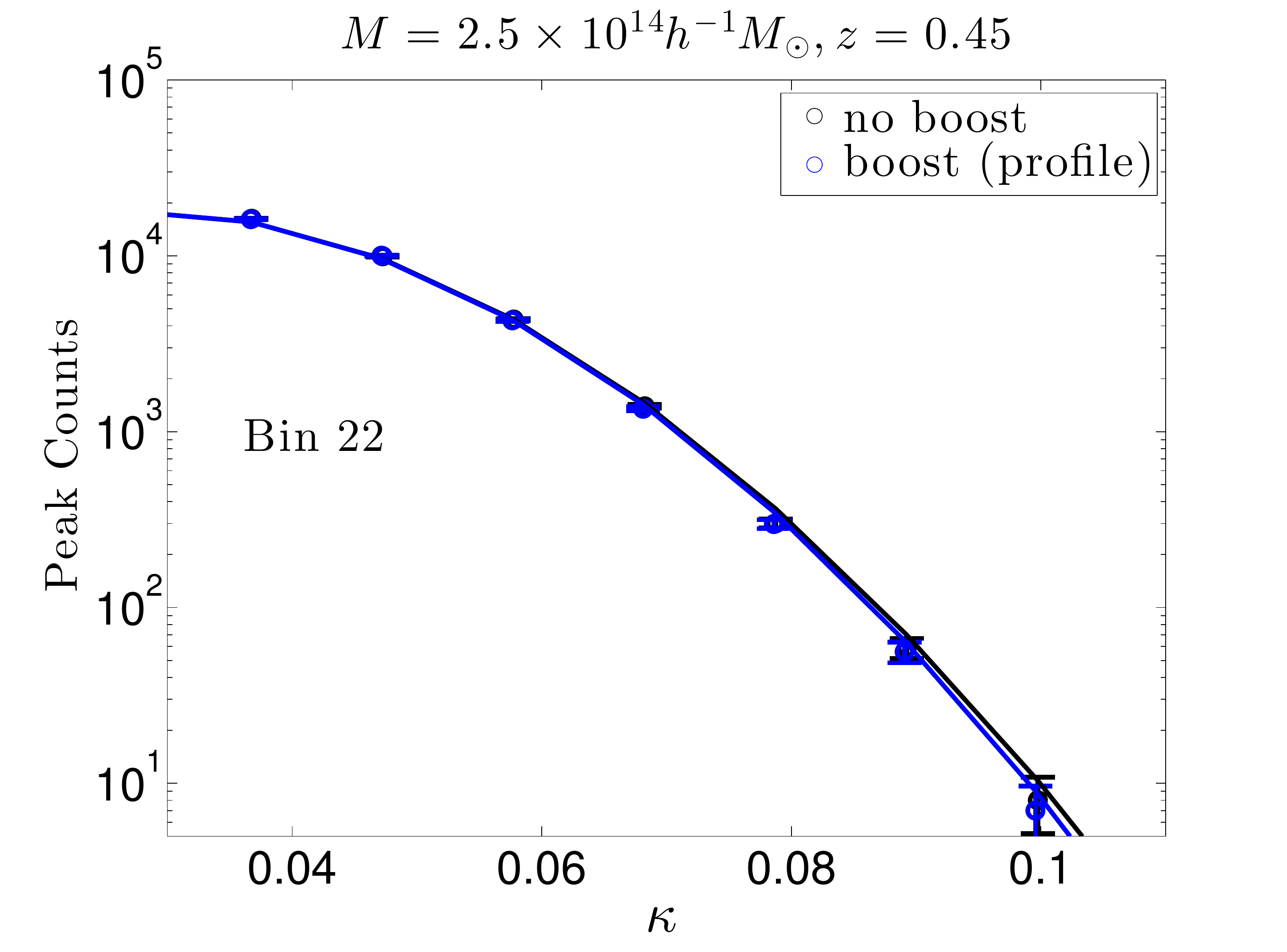}
\includegraphics[width=0.33\textwidth]{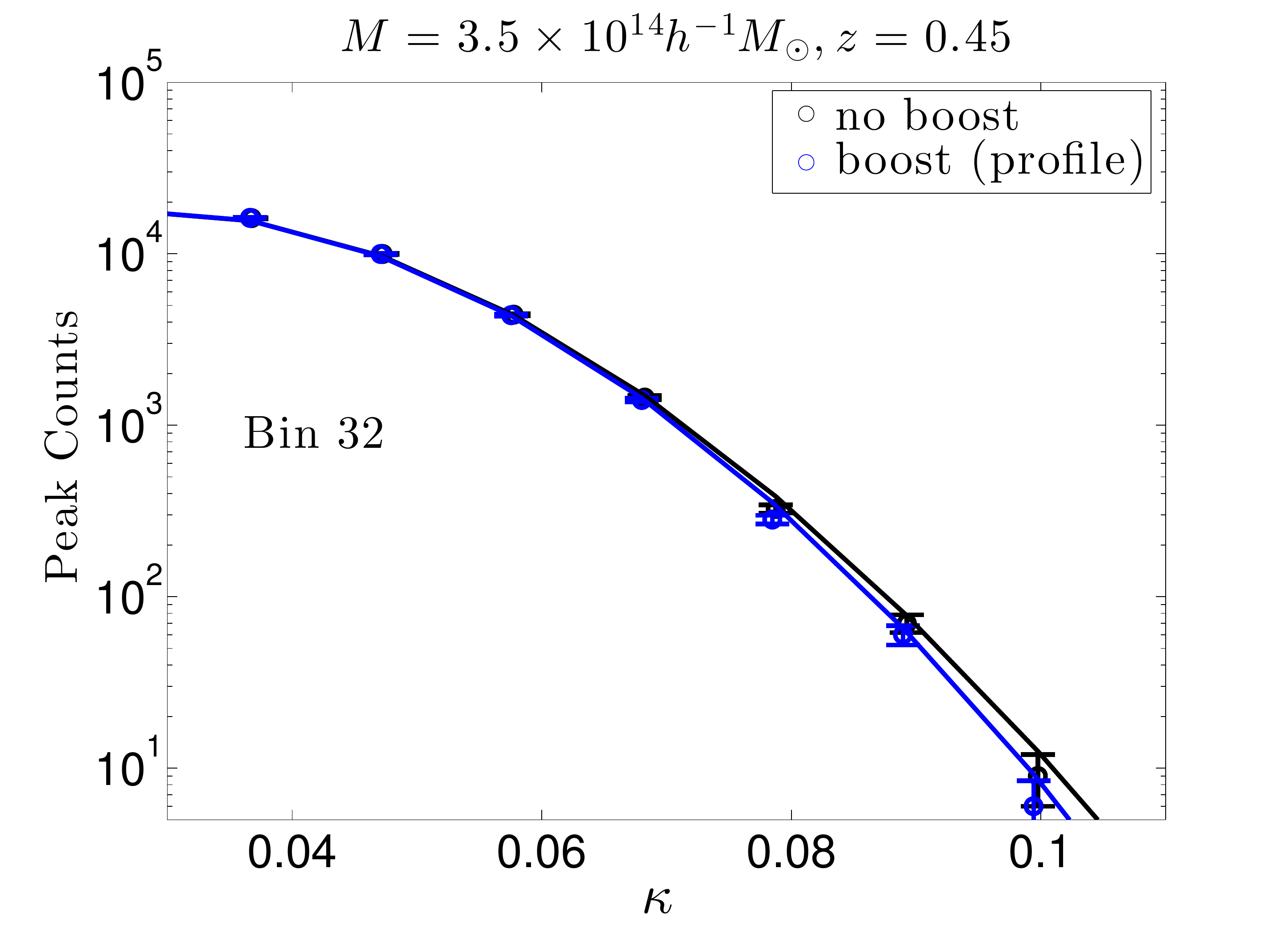}
\caption{The peak distribution with and without the boost factor effects for different bins. The circles
with error bars are the measurement from the mock analysis. The lines with different colour are the
corresponding analytical predictions.}
\label{fig:boost-peak}
\end{figure*}

The cluster member contamination of WL source galaxies depends on the mass and redshift of the cluster.
In order to quantify the boost effect that this has on WL peak counts, we therefore should first analyse
the mass and redshift distribution of clusters that are responsible for the high-SNR peaks in KiDS-450. 

In the noiseless case, a cluster with a given mass and redshift produces a WL peak with a height that can be well 
predicted given the source redshift distribution (see, e.g., Hamana et al. 2004). Fig.~D1 shows the fractional 
contributions of clusters, with different masses and redshifts, to WL peaks of different heights in KiDS. 
It is seen that, without considering the shape noise, the high-SNR peaks correspond to 
clusters with masses larger than $3\times 10^{14} \hbox{ M}_{\odot}$ and in the redshift range up 
to $z\sim 0.4$. 

Taking into account the shape noise, the peak height from a cluster with a given mass and redshift becomes a 
probability function, whose width is dependent on the noise level. Furthermore, noise peaks can occur in halo 
regions, further polluting the sample. The effect of the addition of noise to this sample is shown in Fig.~D2, 
whereby the WL peaks becomes significantly broader in both mass and redshift than that of noiseless case. 
In the figure, we can see that the contributions to the high SNR peaks in our analysis are mainly from the 
massive DM haloes with $M\sim1-5\times10^{14}h^{-1}{\rm M}_{\odot}$ and $z\sim0.1-0.6$.

While, ideally, we would like to have the precise mass and redshift dependence of cluster member contaminations, 
this is impractical given the limited number of cluster candidates. Instead, we opt to divide the KiDS cluster 
candidates (see Radovich et al. 2017) into $6$ bins (see Table~D1). We then use these bins to 
extract the corresponding boost factors, by estimating the excess filling factor (galaxy number over-density) 
distribution around cluster candidates. Fig.~D3 shows the excess galaxy number density (filling factor) 
distributions around cluster candidates for each bin.

\begin{table}
\caption{The cluster samples in six mass and redshift bins used in the boost factor measurement.}
\label{tab:cluster}
\begin{center}
\leavevmode
\begin{tabular}{c c c c} \hline
bin & mass range & $z_{\rm B}$ range & dilution factor\\
\hline
bin11 & $1\leq M/10^{14}{\rm M}_{\odot}/h<2$ & $z_{\rm B}<0.35$ & $1/1.067$ \\
bin12 & $1\leq M/10^{14}{\rm M}_{\odot}/h<2$ & $z_{\rm B}\geq0.35$ & $1/1.108$ \\
bin21 & $2\leq M/10^{14}{\rm M}_{\odot}/h<3$ & $z_{\rm B}<0.35$ & $1/1.135$ \\
bin22 & $2\leq M/10^{14}{\rm M}_{\odot}/h<3$ & $z_{\rm B}\geq0.35$ & $1/1.164$ \\
bin31 & $3\leq M/10^{14}{\rm M}_{\odot}/h<4$ & $z_{\rm B}<0.35$ & $1/1.259$ \\
bin32 & $3\leq M/10^{14}{\rm M}_{\odot}/h<4$ & $z_{\rm B}\geq0.35$ & $1/1.254$ \\
\hline
\end{tabular}
\end{center}
\end{table}

To analyse how the boost factor affects the WL convergence peaks in halo regions, we build an appropriate set 
of mocks. For each of the $6$ bins, we pick out a typical halo with the mass and the redshift as indicated 
in Fig.~D4. We model the halo with the NFW profile, and put it in the centre of a $1.2\times1.2~\rm deg^2$ 
field. We then distribute source galaxies in the field in two ways:
\begin{itemize}
\item[-] no boost: we sample galaxies using our standard KiDS $n_{\rm g}$ and DIR redshift distributions, with 
random intrinsic ellipticity and reduced shear from the central DM halo;
\item[-] boost: based on the no boost case, we further resample member galaxies following the excess galaxy 
number density profile in Fig.~D3. Only random intrinsic ellipticities are given to the member galaxies because 
lensing signals from their own halo should be zero (e.g. Sif\'on et al. 2015).
\end{itemize}
Using the same method, we also generate source catalogs with intrinsic ellipticities set to be zero to produce 
the noiseless cases. Using these galaxy mocks, we then follow the same procedures as done for our KiDS analysis 
to reconstruct the convergence field. Furthermore, we exclude the outermost $\sim 0.1~\rm deg$ regions along each 
side of the field to suppress the boundary effects. For each halo, we do $1000$ realisations according 
to the positions and intrinsic ellipticity distribution of source galaxies. Finally, as the boost effects 
can influence both the WL signal and the noise level in halo regions, we consider them separately with the mocks.

To estimate the WL peak signal, we estimate the ratio of the convergence value of the central peak between the 
two cases with and without the boost effects from the $1000$ noiseless realisations for each halo. We find that 
the average dilution factors for the $6$ bins (see Table~D1). We further test whether a constant boost 
factor, corresponding to the dilution effects in Table~D1, can mimic the real boost effect with radial profiles. 
We do this by resampling member galaxies according to a constant boost factor, such as $1.067$ for the 
case of $\rm bin11$. We find that such a constant boost does indeed model the true boost effect on the 
WL convergence peaks well. Thus in the model calculations, we adopt the constant dilution factors in Table~D1 
in the corresponding $6$ bins.   

For the shape noise levels, we consider the halo and field regions individually. We first calculate a global 
average source number density $n_{\rm g}$ from the data, which includes the excess number density from galaxies 
in clusters. Compared to this global average, the number density in haloes regions $n_{\rm g}^{\rm halo}$ is 
higher depending on the boost factor shown in Fig.~D3; and thus the noise level $\sigma_0^{\rm halo}$ in the 
halo regions is lower. Correspondingly, the number density in field regions $n_{\rm g}^{\rm field}$ is lower 
than $n_{\rm g}$, and $\sigma_0^{\rm field}$ is higher than $\sigma_0$. The three number densities are related by 
\begin{equation}
n_{\rm g}S_{\rm eff}=\sum {n_{\rm g}^{\rm halo}S_{\rm eff}^{\rm halo}} + n_{\rm g}^{\rm field}S_{\rm eff}^{\rm field},
\end{equation}
where $S_{\rm eff}$, $S_{\rm eff}^{\rm halo}$ and $S_{\rm eff}^{\rm field}$ are the total effective area, the 
area occupied by haloes, and the left-over field area with 
$S_{\rm eff}^{\rm field}=S_{\rm eff}-\sum S_{\rm eff}^{\rm halo}$ respectively. From $n_{\rm g}^{\rm field}$,
we can calculate the noise level $\sigma_0^{\rm field}$. It is noted that $\sum S_{\rm eff}^{\rm halo}$, and 
thus also $S_{\rm eff}^{\rm field}$, are cosmological model dependent. 

Using the above equations, we are able to modify our model calculations to include the boost effect as follows: 
\begin{itemize}
\item[(i)] Using Eq.~(4) to calculate peaks in halo regions, we divide the halo mass and redshift into the $6$ 
bins described above. In each bin we include the corresponding constant dilution factor, which acts to modify 
the convergence field from the halo. We also modify the noise level according to the average number density of 
source galaxies in the halo regions. 
\item[(ii)] Using Eq.~(7) to calculate peaks in field regions, we modify the noise level according to Eq.~(D1).
\end{itemize}
These modifications are adopted in our fiducial analysis presented in Sect.~5 including the boost effects.  

To test the model performance, Fig.~D4 shows the peak counts from our $1000$ mock realisations in each of the 
$6$ bins. The blue and black symbols are the peak counts for the cases with and without the boost effects 
respectively. The corresponding solid lines are our model predictions, which demonstrate a good agreement with 
the data in both cases. 

\end{document}